\RequirePackage{silence}
\PassOptionsToPackage{pdfpagelabels=false}{hyperref}
\documentclass[fleqn,usenatbib,useAMS]{mnras} 
\usepackage{graphicx}
\usepackage{amsmath}

\DeclareMathOperator{\sech}{sech}
\usepackage{amsfonts}
\usepackage{float}
\usepackage{bm}
\usepackage[perpage,symbol*]{footmisc}
\usepackage[dash,dot]{dashundergaps}
\usepackage{ae,aecompl}
\usepackage{array}
\usepackage{soul}
\usepackage{mathtools}
\usepackage{multirow}

\DeclareRobustCommand{\appropto}{\mathrel{\vcenter{
		\offinterlineskip\halign{\hfil$##$\cr 
			\propto\cr\noalign{\kern2pt}\sim\cr\noalign{\kern-2pt}}}}}

\DeclareRobustCommand{\perthousand}{%
	\ifmmode
	\text{\textperthousand}%
	\else
	\textperthousand
	\fi}

\DeclareRobustCommand*{\matr}[1]{\mathbfss{#1}}

\title{Detailed numerical implementation of the wide binary test}

\author[Indranil Banik, Charalambos Pittordis \& Will Sutherland]{\parbox[t]{\textwidth} {Indranil Banik$^{1}$\thanks{Email:
\href{mailto:indranilbanik1992@gmail.com}{indranilbanik1992@gmail.com} (Indranil Banik),\newline $\text{\qquad \qquad} ~~$ \href{mailto:c.pittordis@qmul.ac.uk}{c.pittordis@qmul.ac.uk} (Charalambos Pittordis)}, Charalambos Pittordis$^{2}$ \& Will Sutherland$^{2}$} \vspace{10pt} \\
$^1$Scottish Universities Physics Alliance, University of Saint Andrews, North Haugh, Saint Andrews, Fife, KY16 9SS, UK \\
$^2$School of Physics \& Astronomy, Queen Mary University of London, Mile End Road, London, E1 4NS, UK}

\pubyear{2021}
\begin{document}
\label{firstpage}
\pagerange{\pageref{firstpage}--\pageref{lastpage}}

\maketitle

\begin{abstract}

The observed flat rotation curves of galaxies are among a number of astrophysical phenomena which require a larger acceleration than can be provided by the Newtonian gravity of the detected baryons. The main proposed solutions are additional undetected mass in the form of dark matter, or a low-acceleration modification to Newtonian gravity known as Milgromian dynamics (MOND). It is possible to directly test MOND using wide binary stars in the Solar neighbourhood, as these systems should contain a dynamically insignificant amount of dark matter even if it comprises most of the Galaxy. However, local wide binaries in MOND should orbit each other $\approx 20\%$ faster than in Newtonian dynamics. We describe the detailed plan for how this wide binary test will be conducted, focusing especially on stages with a high numerical cost. The computational costs and memory requirements are estimated for the main stages in the plan. Our overall assessment is that the critically important cost function can be evaluated deterministically at a marginal cost of a few seconds, giving the absolute binomial likelihood of a model. This will allow the cost function to be embedded within a Markov Chain Monte Carlo sampler, or a less expensive gradient descent stage designed to reveal the best-fitting model parameters. Therefore, the wide binary test should be feasible using currently available technology.

\end{abstract}

\begin{keywords}
	gravitation -- binaries: general -- stars: kinematics and dynamics -- solar neighbourhood -- astrometry -- methods: statistical
\end{keywords}

\section{Introduction}

Our location within a flattened structure called the Milky Way has been known since ancient times, but we still lack a clear picture of what our Galaxy really is. In particular, the Galaxy cannot be held together in the same way as the Solar System, whose internal motions conform very well to the Newtonian inverse square gravity law. The main hallmark of this is that the rotation speed of more distant planets is smaller, with the planets nicely tracing a Keplerian rotation curve with circular orbital velocity $v_c \propto r^{-1/2}$, where $r$ is the heliocentric distance. Beyond the extent of the luminous matter in our Galaxy, however, the rotation curve is nearly flat, a phenomenon that is clearly detectable in most disk galaxies \citep[for an early review, see][]{Faber_1979}. These dramatic departures from a Keplerian decline could be due to a halo of dark matter surrounding every disk galaxy \citep{Ostriker_Peebles_1973}. Another possibility is Milgromian dynamics \citep[MOND;][]{Milgrom_1983}. In MOND, an isolated point mass generates a gravitational field that declines only as $1/r$ once the acceleration is far below $a_{_0} = 1.2 \times 10^{-10}$~m/s\textsuperscript{2} \citep{Begeman_1991, Gentile_2011, Lelli_2017}.

Wide binary (WB) systems offer a promising way to test MOND because for a Sun-like star, the predicted deviations from Newtonian dynamics become significant for any tracer object $\ga 5$~kAU away \citep{Beech_2009}. This is only a small fraction of the typical interstellar separation in the Solar neighbourhood of $\approx 1$~pc ($648/\mathrm{\pi} = 206.264$~kAU). The expected MOND signal is a 20\% boost to the circular velocity $v_c$ of WBs in the Solar neighbourhood \citep{Banik_2018_Centauri}. Local WBs thus offer a promising way to test MOND. This wide binary test (WBT) has until recently only been considered theoretically \citep[e.g.][]{Hernandez_2012, Pittordis_2018}. This could soon change thanks to the second data release (DR) of the Gaia mission \citep[GDR2;][]{Gaia_2018} and the subsequent Gaia early DR3 \citep[GEDR3;][]{Gaia_2021}. We expect that the full GDR3 will be available by the time the WBT is attempted, due to the time required to construct the algorithms outlined here (estimated at close to a year).

\subsection{Prior work of direct relevance}
\label{Directly_relevant_prior_works}

As described in \citet{Banik_2018_Centauri}, there already exists a MOND WB orbit library containing the probability distribution with respect to the sky-projected separation $r_{\text{sky}}$ and the scaled relative velocity $\widetilde{v}$ as a function of the WB total system mass $M$, semi-major axis $a$, and eccentricity $e$.\footnote{Generalised definitions of $a$ and $e$ for modified gravity theories were provided in their section 2.3.1.} One of the most critical variables is
\begin{eqnarray}
	\widetilde{v} ~\equiv~ v_{\text{rel}} \div \overbrace{\sqrt{\frac{GM}{r_{\text{sky}}}}}^{\text{Newtonian} \, v_c} \, ,
	\label{v_tilde_definition}
\end{eqnarray}
where $v \equiv \left| \bm{v} \right|$ for any vector $\bm{v}$, and $G$ is the Newtonian gravitational constant.

Another version of the WB orbit library is available using the sky-projected scaled relative velocity $\widetilde{v}_{\text{sky}}$, which is based on only the sky-projected part of $\bm{v}_{\text{rel}}$. The WBT will use this version as radial velocities (RVs) are assumed to be unavailable at the precision required for use in the WBT ($\la 30$~m/s). Indeed, RVs are often not available at all, corresponding to an uncertainty equal to the velocity dispersion in the Solar neighbourhood (Appendix \ref{Assigning_RV}). Where known at the few km/s level or better, they can be used to correct for perspective effects, thereby improving the accuracy of the $\widetilde{\bm{v}}_{\text{sky}}$ determination \citep{Banik_2019_Centauri}. In what follows, we will not consider the 3D $\widetilde{v}$, so any instances of $\widetilde{v}$ are intended as a shorthand for $\widetilde{v}_{\text{sky}}$.

A Newtonian version of the orbit library is also available with $5\times$ higher resolution in $e$, which remains fixed along the orbit $-$ unlike in MOND \citep[see figure 20 of][]{Banik_2018_Centauri}. Nonetheless, the Newtonian library is smaller as the index $M$ is not required. This is because there is no special length scale associated with a certain mass, so the only relevant length scale is $a$. However, in MOND, both $a$ and the MOND radius of the mass $\left( r_{_M} \equiv \sqrt{GM/a_{_0}} \right)$ are relevant. $r_{_M}$ depends on the mass, so $M$ is also relevant to the analysis $-$ despite the effect of $M$ mostly being cancelled out by the scaling in Equation \ref{v_tilde_definition}. The residual mass dependency in MOND arises because a WB with high mass will be asymptotically Newtonian within its MOND radius, while a much lower mass WB with the same separation might be in the MOND regime as its MOND radius is smaller. As a result, at fixed $r_{\text{sky}}$, the predicted distribution of $\widetilde{v}$ depends on $M$.

The main purpose of these WB orbit libraries is to allow efficient exploration of different models for the WB population, in particular different distributions of $a$ and $e$. It is also possible to interpolate between the Newtonian and MOND orbit libraries to see which fits the data best. This is equivalent to adding the gravity law $\alpha_{\text{grav}}$ as a model parameter, with $\alpha_{\text{grav}} = 0$ representing Newtonian gravity and $\alpha_{\text{grav}} = 1$ representing MOND. The algorithm we describe will not be suitable for $\alpha_{\text{grav}}$ much outside the range $0-1$, though going a little outside this range should be possible using some extra numerical safety catches. However, it is not anticipated that $\alpha_{\text{grav}}<0$ or $\gg 1$ \citep{Pittordis_2019}. In particular, their results have disproved the notion that WBs in the Solar neighbourhood do not feel the Galactic external field effect (EFE). The EFE is a natural aspect of MOND \citep{Milgrom_1986} due to its inherent non-linearity, and no self-consistent formulations exist without it that also boost galactic rotation curves in the way required of a modified gravity alternative to cold dark matter. In a MOND context, the EFE is also required by data on WBs, confirming that we do not need to allow a boost to Newtonian gravity greatly exceeding the 20\% predicted by MOND \citep{Banik_2018_Centauri}. Thus, the algorithm need not be rated for $\alpha_{\text{grav}} \gg 1$. It should allow slightly negative values in case $\alpha_{\text{grav}} = 0$ (Newtonian), as observational uncertainties could drag it down further.

All of these libraries took only a modest amount of time to prepare, especially in the Newtonian case (though this was partially offset by the use of higher resolution in $e$ because it remains fixed over time, unlike in MOND). The orbit integration in the MOND libraries uses a force library to map out the gravitational field of a point mass experiencing the EFE appropriate for the Solar neighbourhood of our Galaxy \citep{McMillan_2017}. This force library was found by using the ring library protocol \citep[designed for axisymmetric problems, see][]{Banik_2018_escape} to directly integrate the phantom dark matter distribution, avoiding an iterative grid relaxation stage. The ring libraries took about 12 hours on 4 cores (48 core hours) to map out the Newtonian gravitational field of a ring with unit mass and radius. This was critical to the force library, which took about 100 core hours. Only one force library is needed because all WBs in the Solar neighbourhood feel almost the same Galactic external field, but the library must be stretched and scaled to cover different $M$ \citep{Banik_2018_Centauri}. This force library is visualised in \citep{Banik_2019_spacecraft}. The WB orbit library in MOND exploited this symmetry and was similarly expensive to prepare, though the Newtonian WB orbit library took $\approx 5\times$ less as it considered only one WB mass (arbitrarily fixed at $1.5 \, M_\odot$) and did not need to consider different angles between the orbital pole and the Galactic EFE (towards the Galactic centre).

Only changes to numerical parameters are anticipated in the current project as the external field on the Solar neighbourhood is well known \citep{Klioner_2021}, the interpolating function is well constrained by now to the few percent level in the regime of interest \citep{Lelli_2017}, and only the more computer-friendly quasi-linear formulation of MOND \citep[QUMOND;][]{QUMOND} can be explored without an explosion of complexity, risk, and time for code development. Therefore, we will not consider the older aquadratic Lagrangian formulation \citep[AQUAL;][]{Bekenstein_Milgrom_1984}.

\subsection{The role of close binaries (CBs)}
\label{CB_role}

The major source of contamination to the WBT is CB companions to one or both of the detected stars comprising a WB \citep{Belokurov_2020, Clarke_2020}. One major indication is provided by the distribution of WBs with $\widetilde{v} > 3$, which is too large to arise in any plausible modification to gravity \citep{Banik_2018_Centauri}, especially as the mode in $\widetilde{v}$ is close to the Newtonian expectation of $\approx 1$ \citep{Pittordis_2019}. The $r_{\text{sky}}$ distribution of these high $\widetilde{v}$ systems is similar to that of the overall WB population, both of which show a rapid decline with $r_{\text{sky}}$ (see their table 1). This is completely unexpected if the high $\widetilde{v}$ tail is mostly line of sight (LOS) contamination. Moreover, this tail also declines with respect to $\widetilde{v}_{\text{sky}}$, which is also unexpected since in 2D the area in velocity space per unit $\widetilde{v}_{\text{sky}}$ increases with $\widetilde{v}_{\text{sky}}$.

CB contamination can explain these features because the $r_{\text{sky}}$ distribution of the tail naturally follows that of the WBs. The trend with $\widetilde{v}_{\text{sky}}$ can also be explained if we assume that the distribution over the CB semi-major axis $a_{\text{int}}$ is
\begin{eqnarray}
	P \left( a_{\text{int}} \right) ~\propto~ \frac{1}{a_{\text{int}}} \, .
	\label{P_a_CB}
\end{eqnarray}
This is because conservation of probability implies that the distribution of the CB relative velocity $v$ is related to that of $a_{\text{int}}$ by
\begin{eqnarray}
	P \left( v \right) dv ~=~ P \left( a_{\text{int}} \right) da_{\text{int}} \, ,
\end{eqnarray}
while Keplerian motion implies that
\begin{eqnarray}
	\frac{dv}{da_{\text{int}}} ~=~ -\frac{v}{2 a_{\text{int}}} \, .
\end{eqnarray}
Combining these results gives
\begin{eqnarray}
	P \left( v \right) dv ~=~ \frac{2 a_{\text{int}}}{v} P \left( a_{\text{int}} \right) dv \, .
\end{eqnarray}
Substituting in Equation \ref{P_a_CB} shows that we expect
\begin{eqnarray}
	P \left( v \right) ~\propto~ \frac{1}{v} \, .
\end{eqnarray}
This is similar to the observational result of \citet{Pittordis_2019}. Therefore, we assume that the major part of the effort should be expended on accurately handling CB contamination. LOS contamination should also be considered, but this is simpler as it does not depend on the WB properties. The CB contamination does, since the angle between the CB and WB relative velocities must be considered when assessing the impact on $\widetilde{v}_{\text{sky}}$. Thus, CBs must be handled at a much deeper level of the analysis. LOS contamination can be modelled as just adding to the expected counts in each bin of $\left( r_{\text{sky}}, \widetilde{v}_{\text{sky}} \right)$.

\subsection{Operation costs \& programming choices}
\label{Basic_operations}

The computational time requirement (or `cost') associated with various basic numerical operations is summarized in Table \ref{Basic_operations_cost}. The algorithm is designed to operate in \textsc{fortran} to achieve a reasonable level of efficiency and straightforward parallelisation, which is simply not feasible in most other high-level languages. The relatively low-level control allowed by \textsc{fortran} and the efficiency of array manipulations are likely to be critical. \textsc{fortran} achieves a good compromise between efficiency and readability of the code, with modern compilers also providing some very helpful debugging options.

The algorithm must be compiled without line length limits due to its complexity. It will be compiled with a default real number precision of 8 to improve the accuracy. In addition to these compile time options, the algorithm itself will use double precision variables to further improve the accuracy to 16 significant figures.

The aim is to use only one program file for the primary element. While the whole project needs many files, each program is designed to be self-contained apart from the input data files. This will minimize risks associated with linking algorithms, but will make the main program quite large. Risks will be minimized using subroutines, rather than by keeping parts of the main code in different files. The simultaneous use of multiple program files would also make it extremely difficult to efficiently plan alterations.

\begin{table}
	\centering
	\caption{Estimated runtimes of various commonly used basic numerical operations. For parallel computing, each core will have a private copy of the relevant portion of frequently read or updated arrays. Writing to shared arrays will be much less frequent, and only done once the value to be written is clear.}
	\begin{tabular}{cc}
		\hline
		\multicolumn{2}{c}{Basic operation costs} \\
		\hline
		Operation & Cost (ns) \\
		\hline
		$\pm$, $\times$ & 2 \\
		$\div, \sqrt{}$ & 6 \\
		Rounding & 10 \\
		Trigonometric functions & 60 \\ \hline
		Read from vector & 4 \\
		Read from large array & 6 \\
		Write onto vector & 6 \\
		Write onto large array & 10 \\
		\hline
	\end{tabular}
	\label{Basic_operations_cost}
\end{table}

\subsection{Parallel computing}
\label{Parallel_processing_introduction}

The algorithm will rely on parallel computing with the Open Multi-Processing ($omp$) system to complete the desired computations within a reasonable timeframe. The loops to be run in parallel are \underline{underlined} for each major task. We will use the \emph{collapse(2)} directive to run more than one loop in parallel. It is not envisaged that $>2$ loops will be collapsed in order to avoid parallelisation overheads as the number of iterations already greatly exceeds the number of cores.

We assume that 60 cores will be available simultaneously for the WBT. As a result, parallel processing works best with $\approx 1000$ processes being run in parallel. Since they are in general not equally computationally intensive, a dynamic scheduler will be used. This means that each core receives a process as soon as it has completed the previous process assigned to it. Dynamic scheduling should balance the loads fairly well. Even better balancing can be achieved with a larger number of processes, but this will increase the overheads associated with communicating between cores.

Within the parallel portions, the default setting will be that all variables are \emph{firstprivate}. This means that each core has its own copy of each variable, which will be copied across from the main (`master') core. Some variables will be \emph{private} so they are allocated separately for each core, but the initial value is arbitrary. This avoids unnecessarily copying across the value in the master core for large arrays where the array must be unique to each core but the initial value is irrelevant. A small number of variables need to be \emph{shared}, allowing all cores access. Sharing is usually for the purpose of copying some portion of the shared variable into a private array with smaller rank, reducing communication between cores and speeding up read operations by reducing the dimensionality of the array being read from.

Where a variable is shared to allow all cores write access, such write operations are minimized to avoid communication overheads. In addition, special care is taken to avoid conflicts caused by different processes writing onto the same element of a shared array. It is not envisaged that the \emph{atomic} directive will be used to guarantee the avoidance of `races'. Instead, some sections of code will use the \emph{critical} directive to ensure that only one process is executing the critical section at any given time. For safety, such sections will be given names. The use of \emph{critical} sections will be kept to a minimum as these can disrupt the operation of the algorithm by causing cores to wait for each other. In addition, \emph{critical} sections will be kept near the start or end of a parallel section if possible. This is because the cores do have to wait for each other in any case at these points, so \emph{critical} sections there are less likely to impact the overall efficiency.

\section{Further WB calculations}
\label{Further_WB_calculations}

\subsection{Optimising the WB orbit libraries for the WBT}
\label{Optimising_WB}

The WB orbit library calculations described in \citet{Banik_2018_Centauri} serve as crucial a priori predictions for the WBT. While the essential physics is not expected to be altered for either the Newtonian or the MOND WB libraries, the numerical choices will be altered slightly to maximize the accuracy of the WBT. Gaia WBs within 250 pc cover a wider range in $M$ than originally assumed. Moreover, the importance of low $r_{\text{sky}}$ systems as a control has only recently become clear $-$ these should be little affected by MOND, and can thus help to calibrate the CB parameters. To improve the accuracy of the calculations at low $r_{\text{sky}}$, the resolution in $a$ will be significantly better at low $a$. The resolution scheme in both $M$ and $a$ will be very nearly uniform in log-space, and will cover a slightly wider range than in the work of \citet{Banik_2018_Centauri}.

In addition to changing the range and resolution scheme of the WB libraries in $M$ and $a$, another important change will be in how the results are stored. It will become apparent in Section \ref{Scaling_CB_contamination_pattern_for_rtsky} that the CB and WB libraries can be combined much more efficiently if the WB library is stored over the radial variable
\begin{eqnarray}
	\widetilde{r}_{\text{sky}} ~\equiv~ \frac{r_{\text{sky}}}{a} \, .
\end{eqnarray}
This will also significantly improve the resolution at low $r_{\text{sky}}$, where the contributions mostly come from WBs with low $a$ rather than chance alignments of intrinsically quite rare WBs with large $a$. The WB orbit library must cover out to sufficiently large $\widetilde{r}_{\text{sky}}$. In the Newtonian case, $\widetilde{r}_{\text{sky}} \leq 2$. In MOND, slightly higher values can arise, so an upper limit of at least 2.5 is required.

The preparation cost of the WB libraries and their resulting size will thus be somewhat greater than in the work of \citet{Banik_2018_Centauri}, especially as the results must be stored in plain text format for use in the WBT. We estimate that it will take about one week to prepare the WB libraries on one core, though the work can be split across many cores in principle. The memory load should be about 10 GB in plain text format. Therefore, the preparation of the WB orbit libraries should have an affordable computational cost and memory load. The random access memory required to hold the WB orbit libraries is considered further in Section \ref{Master_core_memory_load} alongside the other major memory loads.

\subsection{Storing index ranges in the WB orbit libraries}
\label{Max_indices_WB}

To optimise the calculations for the WBT, we will record the range of non-zero pixels in $\widetilde{r}_{\text{sky}}$ for different $\left(M, a \right)$ in the MOND library and $a$ in the Newtonian library. For each $\widetilde{r}_{\text{sky}}$ up to and including this index, we will record the maximum non-zero pixel in $\widetilde{v}$. It is not envisaged that all of this information will be needed, since only the range in $\left( \widetilde{r}_{\text{sky}}, \widetilde{v} \right)$ should be required for each $\left( M, a \right)$. But the very low computational cost and memory load justifies the recording of this extra information, which could be useful to unlock further efficiency savings beyond those considered in this plan.

\subsection{The Milgromian stretch factors}
\label{Stretch_factor_determination}

One of the most critical aspects of the WBT is obtaining an accurate WB orbit library for an arbitrary value of the gravity law parameter $\alpha_{\text{grav}}$. This needs to be done as part of the main analysis, as described in Section \ref{Gravity_law_interpolation}. The cases $\alpha_{\text{grav}} = 0$ and $\alpha_{\text{grav}} = 1$ (corresponding to Newtonian and Milgromian gravity, respectively) will be known in advance, and these will serve as the starting point.

The basic idea behind the interpolation in $\alpha_{\text{grav}}$ is to determine the stretch factors in $\widetilde{r}_{\text{sky}}$ and $\widetilde{v}_{\text{sky}}$ (which we call $S_r$ and $S_v$, respectively) that transform the Newtonian $\matr{P} \left( \widetilde{r}_{\text{sky}}, \widetilde{v}_{\text{sky}} \right)$ into the MOND one. We will distinguish the gravity theories using the subscripts $N$ (Newton) and $M$ (MOND). It is not possible to exactly obtain $\matr{P}_M \left( \widetilde{r}_{\text{sky}}, \widetilde{v}_{\text{sky}} \right)$ by applying stretch factors to $\matr{P}_N \left( \widetilde{r}_{\text{sky}}, \widetilde{v}_{\text{sky}} \right)$, so further corrections are required $-$ these will be described later (Section \ref{Gravity_law_interpolation}).

\begin{table}
	\centering
	\caption{Estimated cost of the 2D gradient descent required to find the optimal stretch factors in $\widetilde{r}_{\text{sky}}$ and $\widetilde{v}$ such that applying them to the Newtonian WB distribution mimics the MOND result as closely as possible (Section \ref{Stretch_factor_determination}).}
	\begin{tabular}{cc}
		\hline
		\multicolumn{2}{c}{Stretch factors cost} \\
		\hline
		Variable & Pixels \\
		\hline
		$M$ & 15 \\
		$a$ & 39 \\
		$\gamma$ & 41 \\
		$\widetilde{r}_{\text{sky}}$ & 100 \\
		$\widetilde{v}$ & 142 \\
		Target pixels per input pixel & 5 (average) \\
		Stretches per iteration & 3 (to get gradient in 2D) \\
		Steps to convergence & 30 \\
		Basic element (ns) & 20 \\
		Cores assumed & 4 \\
		\hline
		Estimated time (s) & 770 \\
		\hline
	\end{tabular}
	\label{Stretch_factor_determination_cost}
\end{table}

\begin{table}
	\centering
	\caption{Estimated size of the stretch factor libraries holding the optimal factors by which the Newtonian WB distribution should be stretched along the $\widetilde{r}_{\text{sky}}$ and $\widetilde{v}$ axes to mimic the MOND distribution (Section \ref{Stretch_factor_determination}). The resulting errors (or residuals) will not be stored because the size would likely be too large, so these must be recalculated in the main analysis (Section \ref{Gravity_law_interpolation}).}
	\begin{tabular}{ccc}
		\hline
		\multicolumn{2}{c}{Stretch factors size} \\
		\hline
		Variable & Pixels \\
		\hline
		$M$ & 15 \\
		$a$ & 39 \\
		$\gamma$ & 41 \\
		Libraries & 2 \\
		\hline
		Estimated size & \multirow{2}{*}{0.048} \\
		(million entries) & \\
		\hline
	\end{tabular}
	\label{Stretch_factors_size}
\end{table}

Since the eccentricity oscillates over only a few orbits in MOND \citep{Banik_2018_Centauri}, the WB libraries will be marginalised over $e$ for a grid of values in $\gamma$, where
\begin{eqnarray}
	P \left( e \right) ~=~ 1 + \gamma \left( e - \frac{1}{2} \right) \, .
	\label{gamma_definition}
\end{eqnarray}
Different $\left( M, a \right)$ combinations will be considered separately.\footnote{The Newtonian WB libraries do not depend on $M$, but the MOND libraries do. This is because altering $M$ changes the MOND radius relative to fixed length scales of importance to the WBT, e.g. the range of $r_{\text{sky}}$ that will be used.} Each time, the $\matr{P}_N \left( \widetilde{r}_{\text{sky}}, \widetilde{v}_{\text{sky}} \right)$ will be renormalised to a sum of 1, reallocating probabilities of any crashed ($<50$ AU) or escaped ($>100$ kAU) orbits uniformly onto the remaining allowed orbits. We will then apply a 2D gradient descent algorithm \citep{Fletcher_1963} to minimize the squared sum of residuals between the stretched Newtonian array $\overleftrightarrow{\matr{P}}_N$ and the MOND array $\matr{P}_M$. The estimated computational cost is summarized in Table \ref{Stretch_factor_determination_cost}, while the size of the resulting stretch factor libraries for different $\left(M, a, \gamma \right)$ is summarized in Table \ref{Stretch_factors_size}.

\section{Handling close binary contamination}
\label{Close_binaries}

The main complexity in the WBT is including CB contamination (Section \ref{CB_role}) accurately at an affordable computational cost and memory load. CB contamination could in principle be removed by excluding WBs with a CB companion. This is possible to some extent because a CB would cause an astrometric acceleration, leading to a poor GDR2 astrometric solution \citep{Belokurov_2020}. However, removing more distant CBs would entail a significant amount of follow-up observations. Even then, it is likely that a small amount of contamination would remain due to brown dwarfs with $a_{\text{int}} \approx 50$~AU \citep[section 8.2 of][]{Banik_2018_Centauri}. Thus, it is necessary to model the CB contamination. Follow-up observations provide an opportunity to test if this really is the major source of contamination, and to quantify it.

One major simplification is that the CB can be assumed to follow Newtonian dynamics. This is because a CB with large $a_{\text{int}}$ either:
\begin{enumerate}
	\item has negligible impact on the WB $\widetilde{v}_{\text{sky}}$; or
	\item has an `undetected' star that is quite massive and widely separated from the star whose velocity it is contaminating. In this scenario, it is assumed that the third star would have been detected and the WB removed at the data processing stage.
\end{enumerate}

\subsection{Keplerian velocity library}
\label{Keplerian_velocity_library}

To model the CB contamination, we will first find the probability distribution of the orbital velocity for a test particle on a Keplerian orbit with $G = M = a = 1$ and eccentricity $e_{\text{int}}$. The results can then be scaled to different CB and WB mass ratios and semi-major axes. We will assume that the mass ratio of any binary is mass-independent and uniform, though the latter assumption can be adjusted to e.g. increase the weight on equal mass binaries \citep{Badry_2019_twin}. Possible extensions to the main analysis are discussed in Section \ref{Extensions}.

Each $e_{\text{int}}$ will be handled before moving on to the next $e_{\text{int}}$. In the first step for each $e_{\text{int}}$, we will find the orbital phase angle $\phi \left( t \right)$ forward from pericentre using fourth order Runge-Kutta integration to solve
\begin{eqnarray}
	\dot{\phi} ~=~ \frac{h}{r^2} \, ,
	\label{phi_dot}
\end{eqnarray}
where $h$ is the specific angular momentum, $r$ is the distance from the central mass, and an overdot denotes a time derivative. For Keplerian motion with $GMa = 1$,
\begin{eqnarray}
	h ~&=&~ \sqrt{1 - {e_{\text{int}}}^2} \, , \\
	r ~&=&~ \frac{1 - {e_{\text{int}}}^2}{1 + e_{\text{int}} \cos \phi} \, .
	\label{Shape_equation}
\end{eqnarray}

Since the orbital period is $2\mathrm{\pi}$ independently of $e_{\text{int}}$, we will find $\phi$ at a fixed uniformly spaced grid of times covering from pericentre to apocentre, which occurs when $t = \mathrm{\pi}$. A numerical integration will not reach apocentre at exactly this time. The accuracy will therefore be improved by running integrations forwards from pericentre and backwards from apocentre, with the latter starting at $t = \mathrm{\pi}$. The final estimated $\phi$ at each $t$ is then
\begin{eqnarray}
	\phi ~=~ \phi_{\text{PS}} \left( 1 - \frac{t}{\mathrm{\pi}} \right) \, + \, \phi_{\text{AS}} \left( \frac{t}{\mathrm{\pi}} \right) \, ,
	\label{phi_PS_AS_weighting}
\end{eqnarray}
with $\phi_{\text{PS}}$ and $\phi_{\text{AS}}$ denoting the integrations started at pericentre and apocentre, respectively. $\phi_{\text{PS}}$ and $\phi_{\text{AS}}$ will be recorded at every timestep, including the first and last.

\begin{table}
	\centering
	\caption{Runtime estimate for the Keplerian velocity library (Section \ref{Keplerian_velocity_library}). To maximize accuracy, integrations will be done starting at pericentre and running backwards from apocentre. The value of $\phi$ at each time will be found by interpolating between these, giving more weight to the pericentre start when near pericentre (Equation \ref{phi_PS_AS_weighting}).}
	\begin{tabular}{cc}
		\hline
		\multicolumn{2}{c}{Keplerian velocity library cost} \\
		\hline
		Variable & Pixels \\
		\hline
		Starting points & 2 \\
		$e_{\text{int}}$ & 20 \\
		$t$ & 10000 \\
		Runge-Kutta & 4 \\
		Basic element (ns) & 100 \\
		Cores assumed & 1 \\
		\hline
		Estimated time (s) & 0.16 \\
		\hline
	\end{tabular}
	\label{Keplerian_velocity_cost}
\end{table}

\begin{table}
	\centering
	\caption{Estimated size of the Keplerian velocity library (Section \ref{Keplerian_velocity_library}). Although 10,000 steps are used in the orbital integration, the orbital velocity is available at 10,0001 points because the starting point (pericentre or apocentre) is also considered. This requires the first and last timesteps to have a statistical weight half as much as the others.}
	\begin{tabular}{ccc}
		\hline
		\multicolumn{3}{c}{Keplerian velocity library size} \\
		\hline
		Variable & Range & Pixels \\
		\hline
		$e_{\text{int}}$ & $0 - 0.95$ &  20 \\
		$t$ & $0 - \mathrm{\pi}$ & 10001 \\
		\hline
		Estimated size & \multicolumn{2}{c}{\multirow{2}{*}{0.2}} \\
		(million entries) & \multicolumn{2}{c}{} \\
		\hline
	\end{tabular}
	\label{Keplerian_velocity_size}
\end{table}

Once $\phi \left( t \right)$ is found in this way, we will obtain $r \left( t \right)$ using Equation \ref{Shape_equation}. Relative to the circular orbit velocity of 1, energy conservation implies that
\begin{eqnarray}
	v^2 ~&=&~ 1 + 2 \left( \frac{1}{r} - \frac{1}{\underbrace{a}_1}\right) \\
		~&=&~ \frac{1 + 2 \, e_{\text{int}} \cos \phi + {e_{\text{int}}}^2}{1 - {e_{\text{int}}}^2} \, .
\end{eqnarray}
The maximum velocity is thus
\begin{eqnarray}
	v_{\text{max}} ~=~\sqrt{\frac{1 + e_{\text{int}}}{1 - e_{\text{int}}}} \, .
	\label{v_Kepler_max}
\end{eqnarray}
For a maximum CB eccentricity of $e_{\text{int}} = 0.95$, $v_{\text{max}} < 6.25$.

The cost of preparing the Keplerian velocity library is summarized in Table \ref{Keplerian_velocity_cost}, while the memory requirement is summarized in Table \ref{Keplerian_velocity_size}. Results will be recorded at every timestep used in the orbital integration, including the first and last. For safety, we will set $\phi \left( 0 \right) = 0$ and $\phi \left( \mathrm{\pi} \right) = \mathrm{\pi}$. The high cadence is important at high $e_{\text{int}}$, since Equation \ref{phi_dot} shows that $\dot{\phi}$ becomes very large near pericentre. While an adaptive timestep might be more efficient, the low cost and memory load means this is not worthwhile as the goal is to minimize coding $+$ running time.

\subsection{Projected Keplerian velocity library}
\label{Projected_Keplerian_velocity_library}

Once the Keplerian velocity library has been set up (Section \ref{Keplerian_velocity_library}), we will use it to set up another library holding the probability distribution over $\Delta \widetilde{v}_\parallel$ and $\Delta \widetilde{v}_\perp$, the velocity components parallel and orthogonal, respectively, to some randomly chosen vector within the sky plane. The CB could in principle be at any orientation to the sky plane. We define $\theta$ as the angle between the sky-projected portion of the WB $\bm{v}_{\text{rel}}$ and that of the CB, with $\phi$ defining an azimuthal angle. With this notation (standard spherical polar coordinates aligned with the WB $\bm{v}_{\text{sky}}$), $\Delta \widetilde{v}_\parallel$ receives a contribution of $\cos \theta$ while $\Delta \widetilde{v}_\perp$ receives a contribution of $\sin \theta \cos \phi$, with $\cos \theta$ and $\phi$ having uniform distributions. The cost of simulating a grid over these and related variables is summarized in Table \ref{Projected_Keplerian_velocity_cost}, with the memory requirement summarized in Table \ref{Projected_Keplerian_velocity_size}.

\begin{table}
	\centering
	\caption{Estimated cost of the projected Keplerian velocity library (Section \ref{Projected_Keplerian_velocity_library}). The innermost loop should be over $t$, as the protocol is the same in nearly all cases $-$ additional blocks will be created for the first and last timesteps, with the statistical weight halved. The partially underlined variable indicates a possible parallelisation option.}
	\begin{tabular}{cc}
		\hline
		\multicolumn{2}{c}{Projected Keplerian velocity library cost} \\
		\hline
		Variable & Pixels \\
		\hline
		\dashuline{$e_{\text{int}}$} & 20 \\
		$t$ & 10001 \\
		$\theta$ & 1001 \\
		$\phi$ & 1001 \\
		Basic element (ns) & 30 \\
		Cores assumed & 4 \\
		\hline
		Estimated time (s) & 1500 \\
		\hline
	\end{tabular}
	\label{Projected_Keplerian_velocity_cost}
\end{table}

\begin{table}
	\centering
	\caption{Estimated size of the projected Keplerian velocity library (Section \ref{Projected_Keplerian_velocity_library}).}
	\begin{tabular}{ccc}
		\hline
		\multicolumn{3}{c}{Projected Keplerian velocity library size} \\
		\hline
		Variable & Range & Pixels \\
		\hline
		$e_{\text{int}}$ & $0 - 0.95$ & 20 \\
		$\Delta \widetilde{v}_\parallel$ & $0 - 6.25$ & 125 \\
		$\Delta \widetilde{v}_\perp$ & $0 - 6.25$ & 125 \\
		\hline
		Estimated size & \multicolumn{2}{c}{\multirow{2}{*}{0.3}} \\
		(million entries) & \multicolumn{2}{c}{} \\
		\hline
	\end{tabular}
	\label{Projected_Keplerian_velocity_size}
\end{table}

To find the impact of the CB on $\widetilde{v}_{\text{sky}}$, these velocity perturbations must be appropriately scaled and added to the WB $\bm{v}_{\text{sky}}$, which we describe next. Before scaling, the CB library is hypothecated to the undetected companion having a fraction $q_{\text{int}} = 1$ of the total mass in the CB. This is unrealistic since the undetected star should be less massive than the star we detect as part of the WB, i.e. $q_{\text{int}} \leq 0.5$. Moreover, the unscaled CB library is hypothecated to the contaminated star comprising a fraction $q_{\text{ext}} = 1$ of the total detected mass in the WB. This is also unrealistic, since for the other star in the WB to be detected, it should comprise $\ga 0.05$ of the detected mass, i.e. $0.05 \leq q_{\text{ext}} \leq 0.95$.

\subsection{The close binary template (CBT)}
\label{CB_template}

The CBT will hold $P \left( \Delta \widetilde{v}_\parallel, \Delta \widetilde{v}_\perp, \Delta \widetilde{M} \right)$ marginalised over all $q_{\text{int}}$ for the case that only one of the WB stars is contaminated by a CB, the contaminated star makes up a fraction $q_{\text{ext}} = 1$ of the total WB mass, and the CB has $a_{\text{int}} = r_{\text{sky}}$, where $r_{\text{sky}}$ refers to the WB and $a_{\text{int}}$ to the CB. The undetected CB raises the mass of the WB by a fraction $\Delta \widetilde{M}$ above what would be inferred by observers based on the luminosities of the detected stars, one of which is slightly affected by the CB. Since CBs with a smaller separation are expected to be more common (Equation \ref{P_a_CB}), we neglect the positional photocentre-barycentre offset of the CB and assume that its angular separation is so small that light from the undetected star is completely blended with that of the detected star in the WB. Even with accurate spectra, this blended light may be a very small fraction of the total such that the contamination is not readily apparent, especially as the availability of such spectra is not envisaged for the WBT. However, the CB could still have a non-negligible impact on the kinematics and total system mass.

The undetected star induces a recoil on the contaminated star $\propto q_{\text{int}}$, the fraction of the CB mass in the undetected companion. It also raises the CB total mass $\propto \frac{1}{1 - q_{\text{int}}}$. However, we should also consider the small amount of light emitted by the undetected star. When the luminosity is converted to a mass (Appendix~\ref{Mass_luminosity_conversion}), the estimated CB mass will be higher than just that of the detected component. To help the discussion, we define
\begin{eqnarray}
	x ~\equiv~ \frac{\text{Actual CB mass}}{\text{Estimated CB mass}} \, .
	\label{Blended_light_factor}
\end{eqnarray}
Due to the steep mass-luminosity relation \citep[$L \appropto M^{5.5}$ in the Gaia band, equation 4.3 of][]{Pittordis_thesis}, $x > 1$. In fact, the estimated CB mass is expected to be rather close to that of the detected component alone, i.e.
\begin{eqnarray}
	x ~\approx~ \frac{1}{1 - q_{\text{int}}} \, .
	\label{x_estimate}
\end{eqnarray}
Even in the case $q_{\text{int}} = 1/2$, we get that $x = 1.69$ instead of the above estimate of 2 assuming no blended light, for a detected star of actual mass $1 \, M_\odot$. Thus, we approximate that the true mass ratio between the detected components of the WB is almost the same as that inferred from the luminosities, even in the presence of blended light. This remains quite accurate even if the detected components and the undetected star all have the same mass of $1 \, M_\odot$, in which case the WB appears to have a mass ratio of 46:54 rather than the true ratio of 33:67. In addition, if the mass-luminosity relation is a power law, there is no special mass scale, so the calculation of $x$ depends only on the mass ratios and not the actual masses. This is not exactly true in reality, so for simplicity, we will consider a CB where the detected star has mass $1 \, M_\odot$ while its undetected companion has mass $q_{\text{int}} M_\odot/\left(1 - q_{\text{int}}\right)$.

Using these approximations, the fractional increase to the WB mass is
\begin{eqnarray}
	\Delta \widetilde{M} ~=~ q_{\text{ext}} \left( x - 1 \right) \, .
	\label{q_scaling_M}
\end{eqnarray}
All else being equal, a more massive WB has a larger orbital velocity. The hidden mass effect therefore raises the observationally inferred WB $\widetilde{v}$ as only the detected mass enters into Equation \ref{v_tilde_definition} but the actual mass enters into the dynamics. In the Newtonian regime, the boost factor to $\widetilde{\bm{v}}$ is $\sqrt{1 + \Delta \widetilde{M}}$. This factor is also appropriate for WBs with a large separation such that the Galactic EFE dominates, since the WB gravity law is also inverse square in this case. We therefore assume that the same scaling can be applied to the WB $\widetilde{\bm{v}}$ regardless of the gravity law.
\begin{eqnarray}
	\widetilde{\bm{v}} ~\to~ \widetilde{\bm{v}} \sqrt{1 + \Delta \widetilde{M}} \, .
	\label{Hidden_mass_effect}
\end{eqnarray}

The recoil effect on $\widetilde{\bm{v}}$ is more complicated because the apparent photocentre of the detected WB component moves as a result of the CB orbit. In analogy with $q_{\text{int}}$, we define $l_{\text{int}}$ as the fraction of the CB luminosity arising from the `undetected' star. Of the CB relative orbital velocity, the observable fraction is thus $\left| q_{\text{int}} \left( 1 - l_{\text{int}} \right) - \left( 1 - q_{\text{int}} \right) l_{\text{int}} \right|$. With a super-linear mass-luminosity relation, the absolute value clause is in principle unnecessary, but kept for safety. We also need to account for the CB mass exceeding the apparent mass of the detected star by a factor $x$, and that the detected WB component only comprises a fraction $q_{\text{ext}}$ of the total mass of the detected WB stars. Bearing this in mind, the perturbation to $\widetilde{v}$ is
\begin{eqnarray}
	\Delta \widetilde{\bm{v}} ~\propto~ \left| q_{\text{int}} \left( 1 - l_{\text{int}} \right) - \left( 1 - q_{\text{int}} \right) l_{\text{int}} \right| \sqrt{q_{\text{ext}} x}\, .
	\label{q_scaling_v}
\end{eqnarray}

$q_{\text{int}}$ will be marginalised over using a uniform distribution but neglecting the case that $q_{\text{int}} = 1/2$ as then $l_{\text{int}} = 1/2$, making $\Delta \widetilde{\bm{v}}$ vanish. Appendix \ref{Equal_mass_CBs} describes how this special case will be handled as $\Delta \widetilde{M}$ still needs to be considered. In addition to $q_{\text{int}}$, the actual perturbation $\Delta \widetilde{\bm{v}}$ is also $\propto \sqrt{r_{\text{sky}}/a_{\text{int}}}$. In the CBT, $q_{\text{ext}}$ and $r_{\text{sky}}/a_{\text{int}}$ are assumed to be 1. These factors must be put in at a later stage.

\begin{table}
	\centering
	\caption{Estimated cost of the CBT marginalised over $q_{\text{int}}$ (Section \ref{CB_template}).}
	\begin{tabular}{cc}
		\hline
		\multicolumn{2}{c}{CBT cost} \\
		\hline
		Variable & Pixels \\
		\hline
		$e_{\text{int}}$ & 20 \\
		$\Delta \widetilde{v}_\parallel$ & 125 \\
		$\Delta \widetilde{v}_\perp$ & 125 \\
		$q_{\text{int}}$ & 499 \\
		Basic element (ns) & 50 \\
		Cores assumed & 1 \\
		\hline
		Estimated time (s) & 8 \\
		\hline
	\end{tabular}
	\label{CB_template_cost}
\end{table}

\begin{table}
	\centering
	\caption{Estimated size of the CBT (Section \ref{CB_template}). The possible ranges of all variables have been reduced due to the contribution of blended light, which provides particularly significant savings for the velocity perturbations (Equation \ref{q_scaling_v}). This is because low values of $q_{\text{int}}$ cause only a small recoil on the contaminated star, while high values approach the limit that the photocentre and barycentre coincide, the latter moving only due to the WB orbit.}
	\begin{tabular}{ccc}
		\hline
		\multicolumn{3}{c}{CBT size} \\
		\hline
		Variable & Range & Pixels \\
		\hline
		$e_{\text{int}}$ & $0 - 0.95$ & 20 \\
		$\Delta \widetilde{v}_\parallel$ & $0 - 2.65$ & 53 \\
		$\Delta \widetilde{v}_\perp$ & $0 - 2.65$ & 53 \\
		$\Delta \widetilde{M}$ & $\left(-0.05, 0.7 \right)$ & 15 \\
		\hline
		Estimated size & \multicolumn{2}{c}{\multirow{2}{*}{0.8}} \\
		(million entries) & \multicolumn{2}{c}{} \\
		\hline
	\end{tabular}
	\label{CB_template_size}
\end{table}

The estimated cost of preparing the CBT is shown in Table \ref{CB_template_cost}. While the determination of mass from luminosity and vice versa (Appendix \ref{Mass_luminosity_conversion}) is somewhat expensive, this need be done only if $q_{\text{int}}$ is varied, so the basic element is not particularly expensive. The memory load of the resulting arrays is summarized in Table \ref{CB_template_size}. For simplicity, results for different $e_{\text{int}}$ will be stored in separate files.

\subsection{One close binary (1CB) library}
\label{One_CB_library}

We first consider the case where only one of the stars in a WB is contaminated by an undetected CB companion. In this case, we need to marginalise the CBT over different values of $q_{\text{ext}}$ and $a_{\text{int}}/r_{\text{sky}}$. The ultimate goal is to estimate the perturbations caused to a WB with known semi-major axis $a$ and $\widetilde{r}_{\text{sky}} \equiv r_{\text{sky}}/a$. Importantly, the CB is assumed to have a uniform logarithmic prior on $a_{\text{int}}/a$ over a range of 3 dex, ending at some fixed value of $C_{\text{limit}} \equiv a_{\text{int, max}}/a \approx 1/3$ that marks the limit beyond which the CB would become chaotic and not survive over the long term. The 1CB library will be hypothecated to $\widetilde{r}_{\text{sky}} = 1$ and $C_{\text{limit}} = 1$. In the main analysis, it will be necessary to scale the velocity perturbations by $\sqrt{\widetilde{r}_{\text{sky}}/C_{\text{limit}}}$ (Equation \ref{CB_stretch_factor}). The marginalisation over $a_1 \equiv a_{\text{int}}/r_{\text{sky}}$ thus involves considering a grid in $\log_{10} a_1$ with equally weighted values, e.g. weights of $1/6$ on $\left( -2.75, -2.25, -1.75, -1.25, -0.75, -0.25 \right)$. In practice, 70 different values of $a_1$ will be considered, making the logarithmic spacing only $3/70 = 0.043$ dex, or 10\%.

Equation \ref{v_tilde_definition} shows that the dependency of $\Delta \widetilde{\bm{v}}$ on the WB mass ratio is simply $\sqrt{q_{\text{ext}}}$ because the larger the fraction of the detected WB mass in the contaminated star, the larger the CB total mass, increasing its orbital velocity $\propto \sqrt{q_{\text{ext}}}$. However, the Newtonian-predicted orbital velocity of the WB remains the same, since this is based on only the directly observed WB mass. The scaling with $\sqrt{q_{\text{ext}}}$ is also apparent in Equation \ref{q_scaling_v}. Results will be stored separately for different $q_{\text{ext}}$, with the marginalisation over this handled later (Section \ref{Marginalise_CB_libraries}). This allows the use of a different prior on $q_{\text{ext}}$, which is envisaged in an extension to the nominal analysis (Appendix \ref{Equal_mass_CBs}).

To summarize, the 1CB library will be prepared by scaling $\Delta \widetilde{\bm{v}}$ from the CBT (Section \ref{CB_template}) according to
\begin{eqnarray}
	\Delta \widetilde{\bm{v}} ~\to~ \Delta \widetilde{\bm{v}} \sqrt{\frac{q_{\text{ext}}}{a_1}} \, , \quad a_1 \equiv \frac{a_{\text{int}}}{r_{\text{sky}}} \, .
	\label{v_scaling_1CB}
\end{eqnarray}
The library is therefore hypothecated to the case that $a_{\text{int, max}} = r_{\text{sky}}$ of the WB. Since these quantities are in general unequal, the 1CB library must be scaled for use in the WBT, as discussed later in Section \ref{Scaling_CB_contamination_pattern_for_rtsky}.

Since the impact on the total WB mass scales with $q_{\text{ext}}$, the scaling for $\Delta \widetilde{M}$ in the CBT is simply
\begin{eqnarray}
	\Delta \widetilde{M} ~\to~ q_{\text{ext}} \Delta \widetilde{M} \, .
	\label{M_scaling_1CB}
\end{eqnarray}
To implement this, we will first find which bins in $\Delta \widetilde{M}$ receive a non-zero contribution from any given input bin in $\Delta \widetilde{M}$ from the CBT. For each of these output bins, we will find the fraction of the input bin which should be assigned to it. This overlap fraction will be applied to the input probability before applying it as an increment to the appropriate output pixel.

The results will be stored in two libraries, with the inner 1CB library covering a range in $\Delta \widetilde{v}_{\parallel, \perp}$ of $0-20$, while the outer 1CB library will cover $0-90$ at a $4\times$ lower resolution in each parameter. These libraries will be prepared jointly. For any individual contribution, we will first check if the inner library should be incremented, since in this very likely scenario, both libraries should be incremented. Otherwise, we will check if at least the outer library should be incremented.

\begin{table}
	\centering
	\caption{Estimated cost of preparing the inner and outer 1CB libraries (Section \ref{One_CB_library}). Some savings are anticipated from the input range of $\Delta \widetilde{\bm{v}}$ being narrower at low $\Delta \widetilde{M}$, which corresponds to low $q_{\text{int}}$. Note that the range in $\Delta \widetilde{\bm{v}}$ peaks and then declines because for very high $q_{\text{int}}$ and $\Delta \widetilde{M}$, blended light from the undetected star reduces the photocentre-barycentre offset, which vanishes in the case of an equal mass CB (Equation \ref{q_scaling_v}).}
	\begin{tabular}{cc}
		\hline
		\multicolumn{2}{c}{1CB library cost} \\
		\hline
		Variable & Pixels \\
		\hline
		$\gamma$ & 41 \\
		$q_{\text{ext}}$ & 19 \\
		$a_{\text{int}}/r_{\text{sky}}$ & 70 \\
		$\Delta \widetilde{M}$ & 15 \\
		Output bins in $\Delta \widetilde{M}$ & 2 (maximum) \\
		$\Delta \widetilde{v}_\parallel$ & 53 \\
		$\Delta \widetilde{v}_\perp$ & 53 \\
		Efficiency gain (low $\Delta \widetilde{\bm{v}}$ & \multirow{2}{*}{0.6} \\
		range at low $\Delta \widetilde{M}$) & \\
		Basic element (ns) & 30 \\
		Cores assumed & 1 \\
		\hline
		Estimated time (s) & 83 \\
		\hline
	\end{tabular}
	\label{One_CB_cost}
\end{table}

\begin{table}
	\centering
	\caption{Estimated size of the inner (outer) 1CB library (Section \ref{One_CB_library}) without marginalising over $q_{\text{ext}}$. These libraries will be split across files according to the underlined variables.}
	\begin{tabular}{ccc}
		\hline
		\multicolumn{3}{c}{Inner (outer) 1CB library size} \\
		\hline
		Variable & Range & Pixels \\
		\hline
		\underline{$\gamma$} & $\left(-2, 2 \right)$ & 41 \\
		\underline{$q_{\text{ext}}$} & $0.05-0.95$ & 19 \\ [3pt]
		$\Delta \widetilde{M}$ & $\left(-0.05, 0.7 \right)$ & 15 \\
		$\Delta \widetilde{v}_\parallel$ & $0-20$ ($0-90$) & 400 (450) \\
		$\Delta \widetilde{v}_\perp$ & $0-20$ ($0-90$) & 400 (450) \\
		\hline
		Number of files & \multicolumn{2}{c}{779 (779)} \\
		Estimated size & \multicolumn{2}{c}{\multirow{2}{*}{1870 (2370)}} \\
		(million entries) & \multicolumn{2}{c}{} \\
		File size & \multicolumn{2}{c}{\multirow{2}{*}{2.4 (3.0)}} \\
		(million entries) & \multicolumn{2}{c}{} \\
		\hline
	\end{tabular}
	\label{One_CB_size}
\end{table}

The estimated cost of preparing the 1CB library is summarized in Table \ref{One_CB_cost}, while its estimated size is summarized in Table \ref{One_CB_size}. The inner and outer libraries will be stored in separate files, as will results for different $\gamma$ and $q_{\text{ext}}$.

\subsection{Two close binaries (2CB) library}
\label{Two_CB_library}

The most expensive library will almost certainly be that used for handling WBs in which both stars are part of an undetected CB, but the total velocity perturbation is small enough or sufficiently well aligned with the LOS that the WB remains part of the analysis. If the individual star contamination fraction $f_{\text{CB}}$ is significant, this scenario must be considered. Recent results indicate that this is certainly the case (Section \ref{CB_role}), so the 2CB library must be prepared in order to reliably conduct the WBT.

The 2CB library requires us to consider all possible combinations of $\left( \Delta \widetilde{v}_\parallel, \Delta \widetilde{v}_\perp, \Delta \widetilde{M} \right)$ for one star against all possible combinations for the other star detected as part of the WB. As described in Section \ref{One_CB_library}, for fixed $\gamma$ and $q_{\text{ext}}$, the conversion from the CBT to the 1CB library depends only on $a_{\text{int}}/r_{\text{sky}}$ (Equation \ref{v_scaling_1CB}). We therefore need to consider this in addition to the parameters in the CBT, and repeat the calculations for different $\left( \gamma, q_{\text{ext}} \right)$.

Both CBs should follow the same $e_{\text{int}}$ distribution, so it should be safe to assume the same value of $\gamma$ for both. Logically, the CB and WB populations should also have the same $\gamma$, but this assumption need not be hard-wired in the algorithm. Equality of the CB and WB eccentricity distributions will be assumed in the nominal analysis.

For each considered value of $q_{\text{ext}}$, the 2CB library will be prepared by an auto-convolution of the CBT scaled to $q_{\text{ext}}$ with another copy of the CBT scaled to $\left( 1 - q_{\text{ext}} \right)$. Since the mass perturbation $\propto q_{\text{ext}}$ (Equation \ref{M_scaling_1CB}), the total $\Delta \widetilde{M}$ is
\begin{eqnarray}
	\Delta \widetilde{M}_t ~=~ \left( \Delta \widetilde{M} \right)_1 q_{\text{ext}} \, + \, \left( \Delta \widetilde{M} \right)_2 \left( 1 - q_{\text{ext}} \right) \, ,
\end{eqnarray}
where the subscripts indicate which copy of the CBT is used. This will be implemented using a highly accurate 2D binning scheme whereby for each possible $\Delta \widetilde{M}_1$, we will find which $\Delta \widetilde{M}_t$ bins receive a non-zero contribution. For each of these $\Delta \widetilde{M}_t$ bins, we will then determine the relevant area in the plane of $\left( \Delta \widetilde{M}_1, \Delta \widetilde{M}_2 \right)$. This is the area between two parallel lines at a slope of $q_{\text{ext}}/\left( 1 - q_{\text{ext}} \right)$. We will then find which $\Delta \widetilde{M}_2$ bins contribute for the $\left( \Delta \widetilde{M}_1, \Delta \widetilde{M}_t \right)$ pair under consideration, and the corresponding overlap fraction. This will be applied to copy 2 of the CBT. Importantly for the overall efficiency, results for different $\Delta \widetilde{M}_2$ will be bunched together because it is not necessary to consider them separately if the output $\Delta \widetilde{M}_t$ bin is the same.

The velocity perturbations are $\propto \sqrt{q_{\text{ext}}/a_{\text{int}}}$ according to Equation \ref{v_scaling_1CB}, so its analogue for the 2CB library is
\begin{eqnarray}
	\Delta \widetilde{v}_{\parallel, \perp} ~=~ \left( \Delta \widetilde{v}_{\parallel, \perp} \right)_1 \sqrt{\frac{q_{\text{ext}}}{a_1}} \, + \, \left( \Delta \widetilde{v}_{\parallel, \perp} \right)_2 \sqrt{\frac{1 - q_{\text{ext}}}{a_2}} \, .
\end{eqnarray}
Due to the high computational cost, only the most basic forward binning protocol will be used whereby only the central value of the input pixel is considered, so the entire probability within it is assigned to only one output pixel.

The estimated cost of preparing the 2CB library is summarized in Table \ref{Two_CB_cost}. The relatively high cost of the basic element arises from the need to consider $\pm$ combinations for both $\Delta \widetilde{v}_\parallel$ and $\Delta \widetilde{v}_\perp$, which in general means that 8 array operations are required to increment both the inner and the outer 2CB libraries. The costs associated with finding the relevant output velocity indices can in principle be spread over all 41 different $\gamma$ values, but this is not envisaged because the required stack memory is also very large, and is almost directly proportional to the number of steps in $\gamma$. Therefore, the calculations are expected to be parallelised over $\gamma$ and $q_{\text{ext}}$. Due to the need to perform so many operations, the basic element is assigned a rather high cost of 250 ns. In addition, the overall very high cost can cause the computer to slow down to reduce the thermal load, which can raise the cost further. It is envisaged that the preparation of the 2CB library will cause the peak computational load for the whole WBT, except perhaps aspects to be handled by observers regarding Gaia data.

\begin{table}
	\centering
	\caption{Estimated cost of preparing the inner and outer 2CB libraries (Section \ref{Two_CB_library}). Some savings are anticipated from the input range of $\Delta \widetilde{\bm{v}}$ being narrower at low $\Delta \widetilde{M}$, which generally corresponds to low $q_{\text{int}}$. As the problem is symmetric with respect to which star is considered star 1, it is only necessary to consider $q_{\text{ext}} \geq 1/2$, so only 10 pixels are required in $q_{\text{ext}}$ rather than 19. The calculations will be split across different institutions according to the boxed variables, and parallelised over the underlined variables. The loops over the latter will both be run in reverse order so calculations with a lower computational cost are handled later, which should help the dynamic parallel scheduler.}
	\begin{tabular}{cc}
		\hline
		\multicolumn{2}{c}{2CB library cost} \\
		\hline
		Variable & Pixels \\
		\hline
		$\boxed{\gamma}$ & 41 \\
		$\boxed{q_{\text{ext}}}$ & 10 \\
		\underline{$a_{\text{int, 1}}/r_{\text{sky}}$} & 70 \\
		$\Delta \widetilde{v}_{\parallel, 1}$ & 53 \\
		$\Delta \widetilde{v}_{\perp, 1}$ & 53 \\		
		\underline{$\Delta \widetilde{M}_1$} & 15 \\
		$\Delta \widetilde{M}_t$ & 10 (average) \\
		$a_{\text{int, 2}}/r_{\text{sky}}$ & 70 \\
		$\Delta \widetilde{v}_{\parallel, 2}$ & 53 \\
		$\Delta \widetilde{v}_{\perp, 2}$ & 53 \\
		Efficiency gain (low $\Delta \widetilde{v}$ & \multirow{2}{*}{0.6} \\
		range at low $\Delta \widetilde{M}$) & \\
		Basic element (ns) & 250 \\
		Cores assumed & \multirow{2}{*}{500} \\
		(including Strasbourg) & \\
		\hline
		Estimated time (d) & 8.3 \\
		\hline
	\end{tabular}
	\label{Two_CB_cost}
\end{table}

The estimated sizes of the inner and outer 2CB libraries are shown in Table \ref{Two_CB_size}. These cover the same range in $\Delta \widetilde{\bm{v}}$ and $\Delta \widetilde{M}$ as the corresponding 1CB libraries, but the range in $q_{\text{ext}}$ is smaller as only $q_{\text{ext}} \geq 1/2$ needs to be considered in the 2CB case. Therefore, the 2CB library will be somewhat smaller than the 1CB library despite the vastly higher computational cost of preparing it. This is fortunate because a large proportion of the files for the 2CB library will need to be transferred between different institutions, which will in principle be able to handle any arbitrary selection of $\left( \gamma, q_{\text{ext}} \right)$ negotiated to avoid substantially repeating calculations.

\begin{table}
	\centering
	\caption{Estimated size of the inner (outer) 2CB library (Section \ref{Two_CB_library}) without marginalising over $q_{\text{ext}}$. These libraries will be split across files according to the underlined variables. The files will need to be combined from different institutions.}
	\begin{tabular}{ccc}
		\hline
		\multicolumn{3}{c}{Inner (outer) 2CB library size} \\
		\hline
		Variable & Range & Pixels \\
		\hline
		\underline{$\gamma$} & $\left(-2, 2 \right)$ & 41 \\
		\underline{$q_{\text{ext}}$} & $0.05-0.95$ & 10 \\ [3pt]
		$\Delta \widetilde{M}$ & $\left(-0.05, 0.7 \right)$ & 15 \\
		$\Delta \widetilde{v}_\parallel$ & $0-20$ ($0-90$) & 400 (450) \\
		$\Delta \widetilde{v}_\perp$ & $0-20$ ($0-90$) & 400 (450) \\
		\hline
		Number of files & \multicolumn{2}{c}{410 (410)} \\
		Estimated size & \multicolumn{2}{c}{\multirow{2}{*}{984 (1245)}} \\
		(million entries) & \multicolumn{2}{c}{} \\
		File size & \multicolumn{2}{c}{\multirow{2}{*}{2.4 (3.0)}} \\
		(million entries) & \multicolumn{2}{c}{} \\
		\hline
	\end{tabular}
	\label{Two_CB_size}
\end{table}

\subsection{Marginalising over $q_{\text{ext}}$}
\label{Marginalise_CB_libraries}

After this rather expensive step, the 2CB library must be marginalised over $q_{\text{ext}}$, with a uniform or sloped distribution assumed for the nominal analysis (an extension is envisaged as described in Appendix \ref{Equal_mass_CBs}). To minimize the random access memory load during the marginalisation, each of the files described in Table \ref{Two_CB_size} will be loaded and added to the combined array before moving on to the next file. The 1CB library will be marginalised over $q_{\text{ext}}$ in the same way. The computational cost of marginalising the CB libraries over $q_{\text{ext}}$ is assumed to be negligible in comparison to the other costs, while the memory load from the CB libraries is also much smaller once they have been marginalised over $q_{\text{ext}}$.

Alterations to the assumed distribution of $q_{\text{int}}$ would require remaking the CBT (Section \ref{CB_template}) and all stages dependent on it. Such alterations would therefore be very expensive, and are not envisaged.

However, we will consider an alteration to the $q_{\text{ext}}$ distribution of the form discussed in Appendix \ref{Equal_mass_CBs}. This alteration is well motivated observationally \citep{Badry_2019_twin}, and can be done at a relatively modest cost as the most expensive libraries do not need to be redone.

\section{Handling observations}
\label{Observations}

The WBT will use a WB catalogue prepared similarly to that in \citet{Pittordis_2019}, but updated to GEDR3 \citep{Gaia_2021} or the full GDR3 if available. The catalogue will cover WBs out to 250 pc with $r_{\text{sky}} = 1.5-40$ kAU and $\widetilde{v}_{\text{sky}} \leq 8$, allowing additional restrictions to be imposed later. The main one imposed before the analysis described here is that both stars must be brighter than apparent magnitude 17 in the Gaia band, since the data quality deteriorates for fainter stars. In addition, WBs will be excluded if in regions of the sky with known star clusters, where LOS contamination would surely be enhanced.

\subsection{Monte Carlo (MC) trials to propagate observational uncertainties}
\label{WB_MC_trials}

To estimate the impact of observational uncertainties, we will conduct $2^{12}$ MC trials for each WB (Section \ref{Sampling_invariant_prior} clarifies why it is best to use a power of 2). Each MC trial will involve using the pre-calculated eigenvalues and eigenvectors of the $5 \times 5$ Gaia covariance matrix for each star, with the variables being its position, parallax, and proper motion. The eigenvalue and eigenvector calculations will take advantage of a highly efficient scheme to diagonalise a $5 \times 5$ symmetric matrix that IB previously developed and had independently tested. Therefore, the cost of diagonalising the covariance matrix for each star is assumed to be negligible in comparison to the other costs.

The basic element of each MC trial will mostly consist of generating 5 Gaussian random numbers, and following through how the resulting perturbations along each eigenvector affect $r_{\text{sky}}$ and $\widetilde{v}$. The shift to the nominal set of parameter values is
\begin{eqnarray}
	\text{\textbf{Error}} ~=~ \sum_{i = 1}^5 G_i \sqrt{\lambda_i} \left( \widehat{\bm{e}}_i \cdot \bm{\sigma} \right) \, ,
\end{eqnarray}
where the $G_i$ are independently distributed Gaussian random numbers with dispersion 1, the $\lambda_i$ are the eigenvalues of the correlation matrix ($\equiv$ the covariance matrix normalised so diagonal elements are 1), $\widehat{\bm{e}}_i$ are the corresponding unit eigenvectors, and $\bm{\sigma}$ is a list of the uncertainties in each parameter if the covariance matrix contained only its diagonal elements. Note that the $\cdot$ indicates element-wise multiplication, the result of which is a vector of length 5.

In addition, the LOS separation must also be estimated using an MC approach. We envisage that this will be done using $r_{\text{sky}}$ in the manner of \citet{Banik_2019_Centauri}. The basic principle is that if the stars in a WB are separated by 10 kAU on the sky, their LOS separation is likely not much higher, allowing this to be guessed using a stochastic approach. But if the relative distance uncertainty from Gaia parallaxes has a higher precision, that will be used instead. Some hybrid method is also possible. However, it is likely that in the vast majority of cases, $r_{\text{sky}}$ is smaller than the Gaia measurement of the difference in trigonometric distances.

The RV of the WB will in general be unknown for the time being, even with the plausible assumption that it is the same for both stars. The process of allocating the systemic RV is described in Appendix \ref{Assigning_RV}. A random number must first be used to decide which Galactic disk component the WB is part of, with the WB rejected if its proper motion is inconsistent with disk kinematics at its position. Another Gaussian random number is then necessary to assign the RV based on the mean and dispersion in RV for a star at that location drawn from the selected disk component. The proper motion, disk membership, and RV can all vary between MC trials, so the total sample size is in general not always the same.

It will be important to propagate uncertainties on the parallaxes and proper motions of both stars before calculating $r_{\text{sky}}$ and $\widetilde{v}$ for the two considered as a WB. Since Gaussian random numbers are most efficiently generated in pairs, we will generate a pair of Gaussian random numbers on alternating steps, which we will keep track of by using the possibility of negating logicals using the $.not.$ operator or its equivalent.

\begin{table}
	\centering
	\caption{Estimated cost of preparing the MC trials for each WB (Section \ref{WB_MC_trials}).}
	\begin{tabular}{cc}
		\hline
		\multicolumn{2}{c}{WB MC trials cost} \\
		\hline
		Variable & Number \\
		\hline
		Wide binaries & 25,000 \\
		Stars per binary & 2 \\
		Monte Carlo trials & $2^{12} = 4096$ \\
		Basic element (ns) & 10,000 \\
		Cores assumed & 1 \\
		\hline
		Estimated time (s) & 2050 \\
		\hline
	\end{tabular}
	\label{WB_MC_trials_cost}
\end{table}

The estimated cost of preparing the MC trials is shown in Table \ref{WB_MC_trials_cost}. The need for several trigonometric function evaluations leads to the high estimated cost of the basic element, which in addition requires several Gaussian random numbers and other expensive calculations like exponentials. There may also be non-negligible overheads for each WB to determine the likelihood of membership to different Galactic disk components.

\subsection{The invariant prior (IP) method}
\label{Invariant_prior_method}

The observational input to the statistical analysis that does the WBT is what we term the `photograph', which holds the number of WBs in different $\left( r_{\text{sky}}, \widetilde{v} \right)$ bins. Table \ref{Pixel_decision_r} shows our decision on which pixels should be used in $r_{\text{sky}}$, while Table \ref{Pixel_decision_v} shows our decision on the pixellation scheme in $\widetilde{v}$. In both cases, the bins are wider at larger values due to the declining probability density, and also because the physics is not expected to change much between e.g. $\widetilde{v} = 4$ and $\widetilde{v} = 5$, whereas the distinction between $\widetilde{v} = 1$ and $\widetilde{v} = 1.2$ is critical to the WBT. All pixels should have $\ga 10$ WBs. It is anticipated that there will be $\approx 6000$ WBs suitable for the WBT in the photographic range of $\left( r_{\text{sky}}, \widetilde{v} \right)$ after additional quality cuts are imposed, so the number of WBs in different pixels should be fairly similar. Binomial uncertainties are expected to be significant for any individual pixel, but the WBT may well be feasible once all 540 pixels are considered.

\begin{table}
	\centering
	\caption{The binning scheme in $r_{\text{sky}}$ for the WBT. There will be 18 bins altogether, covering the range $2-30$ kAU.}
	\begin{tabular}{ccc}
		\hline
		Range (kAU) & Pixel size (kAU) & Pixels \\
		\hline
		$2-5$ & 0.5 & 6 \\
		$5-10$ & 1 & 5 \\
		$10-16$ & 2 & 3 \\
		$16-22$ & 3 & 2 \\
		$22-30$ & 4 & 2 \\
		\hline
		$2-30$ & \multicolumn{2}{c}{Total pixels: 18} \\
		\hline
	\end{tabular}
	\label{Pixel_decision_r}
\end{table}

\begin{table}
	\centering
	\caption{The binning scheme in $\widetilde{v}_{\text{sky}}$ for the WBT. There will be 30 bins altogether, covering the range $0-5$.}
	\begin{tabular}{ccc}
		\hline
		Range & Pixel size & Pixels \\
		\hline
		$0-1.6$ & 0.08 & 20 \\
		$1.6-2.6$ & 0.2 & 5 \\
		$2.6-5$ & 0.48 & 5 \\
		\hline
		$0-5$ & \multicolumn{2}{c}{Total pixels: 30} \\
		\hline
	\end{tabular}
	\label{Pixel_decision_v}
\end{table}

To generate the photograph with the underlying rather than the observed $\left( r_{\text{sky}}, \widetilde{v} \right)$ distribution, we will use the IP method. This is an iterative non-parametric method to recover the underlying probability distribution of some observable given its observed distribution and the measurement uncertainty for each point. Therefore, we assume no particular functional form for the $\left( r_{\text{sky}}, \widetilde{v} \right)$ distribution.

The basic principle is that if e.g. the population is distributed as a Gaussian with mean 0 and dispersion 1, then a point observed at $1.5 \pm 0.5$ was almost certainly scattered up from $\approx 1$. This is overwhelmingly more likely than being scattered down from 2. However, standard MC error propagation does not capture this because the error bars are symmetric. The IP method begins with the $\left( r_{\text{sky}}, \widetilde{v} \right)$ distribution generated from all the MC trials of \emph{all} WBs, which is considered a prior for the next iteration. In the next iteration, the MC trials for each WB are given weights according to the prior likelihood of finding a WB at that point in parameter space. These weights are then renormalised to a sum of $1/N$ for each WB, where $N$ is the total number of WBs. Each MC trial for each WB then contributes towards building up a new population distribution of $\left( r_{\text{sky}}, \widetilde{v} \right)$, which will end up correctly normalised if edge effects can be neglected. The resulting population distribution is in general different to the prior used to build it up, so the new population distribution is again used as a prior for a further iteration. It is hoped that the $\left( r_{\text{sky}}, \widetilde{v} \right)$ distribution will converge after about 10 iterations, as demonstrated in some simplified tests of the IP method.

At each iteration, the MC trials for each WB must be reweighted according to the prior for that stage, which is the population distribution at the previous stage. Importantly for the overall efficiency, \textbf{the MC trials for each WB should not be recalculated, merely reweighted}. Therefore, we assume that the computational cost of the IP iterations is negligible in comparison to the other costs.

\subsubsection{Handling edge effects}
\label{Edge_effects_IP}

The IP must be interpolated to find the relative likelihood of any particular MC trial. This will be done using nearest corner interpolation (Section \ref{Scaling_CB_contamination_pattern_for_rtsky}). To better handle edge effects, this will be done in the space of $\ln P$ rather than $P$. This will make it much easier to handle MC trials outside the photographic range, as a power-law extrapolation will automatically be used.

\subsection{Sampling the IP}
\label{Sampling_invariant_prior}

Once we have found the IP, by definition the posterior population distribution weighted by this prior is the same as the prior. To sample it for one WB, we will first find the relative probabilities of its $2^{12}$ different MC trials and normalise these to a sum of 1, thereby giving the probability that each MC trial is the `right' one for this WB. We will then randomly choose one of these MC trials, with the likelihood of any particular choice matching the above calculation. To improve the efficiency, we will construct a cumulative probability array where e.g. element 97 holds the total probability for MC trials $1-97$. We will then determine at which MC trial the cumulative probability first exceeds a randomly generated number. This stage will use the bisection method, which is why the number of MC trials will be a power of 2. In this way, we will obtain a list of 10,000 $\left( r_{\text{sky}}, \widetilde{v} \right)$ combinations for that particular WB drawn from the convolution of its error bar (found by the MC trials) with the population IP.

A similar approach will be followed for the other WBs. In this way, we will build up 10,000 `photographs' containing the $\left( r_{\text{sky}}, \widetilde{v} \right)$ distribution inferred from observations. Each time, we will also record the total number of WBs within the region of interest ($r_{\text{sky}} = 2-30$~kAU, $\widetilde{v} = 0-5$). These 10,000 photographs will be fixed to have a consistent point of comparison for different models, the generation of which is discussed in Section \ref{Cost_function}.

\begin{table}
	\centering
	\caption{Estimated cost of preparing the observed photographs by sampling the IP (Section \ref{Sampling_invariant_prior}).}
	\begin{tabular}{cc}
		\hline
		\multicolumn{2}{c}{IP sampling cost} \\
		\hline
		Variable & Number \\
		\hline
		Wide binaries & 10,000 \\
		IP samples (photographs) & 10,000 \\
		Steps to select MC trial via bisection & 12 \\
		Basic element (ns) & 100 \\
		Cores assumed & 1 \\
		\hline
		Estimated time (s) & 120 \\
		\hline
	\end{tabular}
	\label{Observed_photographs_cost}
\end{table}

\begin{table}
	\centering
	\caption{Estimated size of the photographs obtained by sampling the IP (Section \ref{Sampling_invariant_prior}). The range of each pixel in $r_{\text{sky}}$ ($\widetilde{v}$) is given in Table \ref{Pixel_decision_r} (\ref{Pixel_decision_v}).}
	\begin{tabular}{ccc}
		\hline
		\multicolumn{3}{c}{Observed photographs size} \\
		\hline
		Variable & Range & Pixels \\
		\hline
		$r_{\text{sky}}$ & $2 - 30$~kAU & 18 \\
		$\widetilde{v}$ & $0 - 5$ & 30 \\
		IP samples & \multicolumn{2}{c}{10000} \\
		\hline
		Estimated size & \multicolumn{2}{c}{\multirow{2}{*}{5.4}} \\
		(million entries) & \multicolumn{2}{c}{} \\
		\hline
	\end{tabular}
	\label{Observed_photographs_size}
\end{table}

Table \ref{Observed_photographs_cost} estimates the cost of obtaining the 10,000 observed photographs by sampling the IP. The estimated size of these photographs is summarized in Table \ref{Observed_photographs_size}.

\section{Comparing theory with observations $-$ the cost function}
\label{Cost_function}

The primary element of the WBT will be a Markov Chain Monte Carlo (MCMC) analysis to infer the gravity law parameter $\alpha_{\text{grav}}$, marginalising over other parameters associated with the problem. The model parameters are summarized in Table \ref{Model_parameter_list}. $P \left( a \right)$ will be modelled using a broken power law with a low-end slope of $-1$ \citep{Opik_1924}, so the break radius $a_{\text{break}}$ and high-end slope $a_{\text{slope}}$ yield two model parameters. $P \left( a_{\text{int}} \right)$ will be modelled as a single power law with slope $-1$ that cannot readily be altered. Linear distributions will be used for $P \left( e \right)$ and $P \left( e_{\text{int}} \right)$, with the nominal analysis assuming the same distribution of $e$ for the CB and WB populations. If significantly different slopes are preferred, there may be something wrong with the model as the CB population should smoothly transition into the WB population. Thus, a single parameter $\gamma$ should capture the binary eccentricity distribution.

\begin{table}
	\centering
	\caption{List of model parameters. The first two are required to specify $P \left( a \right)$ for the WB population, since the low-end logarithmic slope is hard-wired at $-1$ \citep{Opik_1924} and cannot readily be altered $-$ this also applies to the CB population. The best-fitting parameter values and their uncertainties will be inferred using a Markov Chain Monte Carlo process with $10^5$ trials (Section \ref{Cost_function}).}
	\begin{tabular}{ccc}
		\hline
		\multicolumn{3}{c}{Model parameters} \\
		\hline
		Variable & Meaning & Range \\
		\hline
		$a_{\text{break}}$ & Break radius in $P \left( a \right)$ of the WBs & $>1$~kAU \\
		$a_{\text{slope}}$ & High-end logarithmic slope of $P \left( a \right)$ & $<-1$ \\
		$\gamma$ & Distribution of $e$ (Equation \ref{gamma_definition}) & $\left( -2, 2 \right)$ \\
		$f_{\text{CB}}$ & Individual star contamination fraction & $0-1$ \\
		$C_{\text{limit}}$ & Maximum $a_{\text{int}}/a$ for stable CB orbits & $0.1-0.7$ \\
		\multirow{2}{*}{$f_{\text{LOS}}$} & Line of sight contamination fraction & \multirow{2}{*}{$<0.1$} \\
		& for $r_{\text{sky}} = 2-30$~kAU, $\widetilde{v}_{\text{sky}} = 0-5$ & \\
		$\alpha_{\text{grav}}$ & Gravity law (0: Newton; 1: MOND) & $\left( -2, 2.5 \right)$ \\
		\hline
	\end{tabular}
	\label{Model_parameter_list}
\end{table}

The LOS contamination will be modelled as $\propto \widetilde{v}_{\text{sky}} \sqrt{r_{\text{sky}}}$, bearing in mind geometric factors and the narrower range in physical velocity at larger $r_{\text{sky}}$ (Equation \ref{v_tilde_definition}). Since this scaling is very different to the WB population, it is likely that very stringent upper limits will be obtained on $f_{\text{LOS}}$, the LOS contamination fraction. Indeed, \citet{Pittordis_2018} already demonstrated that LOS contamination can explain at best only a very small fraction of the detected WBs in GDR2.

MCMC analyses typically take order 100,000 iterations to yield accurate results. Therefore, it is critically important that the relative likelihood of each model (the `cost function') can be obtained in a few seconds. In the rest of this section, we describe the most expensive stages in the evaluation of the cost function. Its overall estimated cost is discussed in Section \ref{Overall_cost}.

\subsection{Memory load on the master core}
\label{Master_core_memory_load}

To evaluate the cost function, a significant amount of information must be available in the random access memory in the form of a few key libraries. These are too large for each core to store a private copy, so the arrays will be shared. This section estimates the memory requirement created by the largest such arrays.

To handle CB contamination, it is necessary to have the 1CB and 2CB libraries, each of which is split into an inner part and an outer part at lower resolution but covering a much wider range. The memory load from the inner CB libraries is estimated in Table \ref{Inner_CB_size}, while Table \ref{Outer_CB_size} shows the corresponding slightly larger estimate for the outer CB libraries.

\begin{table}
	\centering
	\caption{Estimated memory load on the master core from the inner 1CB and 2CB libraries.}
	\begin{tabular}{cc}
		\hline
		\multicolumn{2}{c}{Inner CB libraries size} \\
		\hline
		Variable & Pixels \\
		\hline
		Libraries & 2 \\
		$\Delta \widetilde{v}_\parallel$ & 400 \\
		$\Delta \widetilde{v}_\perp$ & 400 \\
		$\Delta \widetilde{M}$ & 15 \\
		$\gamma$ & 41 \\
		\hline
		Estimated size & \multirow{2}{*}{197} \\
		(million entries) & \\
		\hline
	\end{tabular}
	\label{Inner_CB_size}
\end{table}

\begin{table}
	\centering
	\caption{Estimated memory load on the master core from the outer 1CB and 2CB libraries.}
	\begin{tabular}{cc}
		\hline
		\multicolumn{2}{c}{Outer CB libraries size} \\
		\hline
		Variable & Pixels \\
		\hline
		Libraries & 2 \\
		$\Delta \widetilde{v}_\parallel$ & 450 \\
		$\Delta \widetilde{v}_\perp$ & 450 \\
		$\Delta \widetilde{M}$ & 15 \\
		$\gamma$ & 41 \\
		\hline
		Estimated size & \multirow{2}{*}{249} \\
		(million entries) & \\
		\hline
	\end{tabular}
	\label{Outer_CB_size}
\end{table}

The WBT also requires WB libraries in the gravity theories under consideration. The estimated size of the Newtonian WB library is shown in Table \ref{Newton_WB_size}, while the corresponding estimate for MOND is shown in Table \ref{MOND_WB_size}. The MOND WB library is significantly larger due to the need to store results for different $M$, though it has $5\times$ lower resolution in $e$.

\begin{table}
	\centering
	\caption{Estimated memory load on the master core from the Newtonian WB library.}
	\begin{tabular}{cc}
		\hline
		\multicolumn{2}{c}{Newtonian WB library size} \\
		\hline
		Variable & Pixels \\
		\hline
		$a$ & 39 \\
		$e$ & 100 \\
		$\widetilde{r}_{\text{sky}}$ & 100 \\
		$\widetilde{v}$ & 145 \\
		\hline
		Estimated size & \multirow{2}{*}{57} \\
		(million entries) & \\
		\hline
	\end{tabular}
	\label{Newton_WB_size}
\end{table}

\begin{table}
	\centering
	\caption{Estimated memory load on the master core from the MOND WB library.}
	\begin{tabular}{cc}
		\hline
		\multicolumn{2}{c}{MOND WB library size} \\
		\hline
		Variable & Pixels \\
		\hline
		$M$ & 15 \\
		$a$ & 39 \\
		$e$ & 20 \\
		$\widetilde{r}_{\text{sky}}$ & 125 \\
		$\widetilde{v}$ & 180 \\
		\hline
		Estimated size & \multirow{2}{*}{263} \\
		(million entries) & \\
		\hline
	\end{tabular}
	\label{MOND_WB_size}
\end{table}

Besides the CB and WB libraries required to calculate the simulated photo, it is also necessary to have observed photo(s) with which to compare it. For the WBT, it is envisaged that 10,000 different versions of the observed photo will be generated using the IP method (Section \ref{Invariant_prior_method}). The resulting memory load on the master core is estimated in Table \ref{Observed_photos_size}.

\begin{table}
	\centering
	\caption{Estimated memory load on the master core from the 10,000 different versions of the observed photograph determined using the IP method (Section \ref{Invariant_prior_method}).}
	\begin{tabular}{cc}
		\hline
		\multicolumn{2}{c}{Observed photographs size} \\
		\hline
		Variable & Pixels \\
		\hline
		Number of photos & 10,000 \\
		$r_{\text{sky}}$ (Table \ref{Pixel_decision_r}) & 18 \\
		$\widetilde{v}$ (Table \ref{Pixel_decision_v}) & 30 \\
		\hline
		Estimated size & \multirow{2}{*}{5.4} \\
		(million entries) & \\
		\hline
	\end{tabular}
	\label{Observed_photos_size}
\end{table}

\subsubsection{Overall memory load on the master core}
\label{Overall_master_core_memory_load}

The overall memory load on the master core is estimated in Table \ref{Master_core_memory_load_summary}, considering only the largest arrays. The total load is about 800 million double precision (16 digit) numbers, which should be manageable on modern computer clusters. However, it would be difficult for individual computers to handle. Therefore, tests of the algorithm will need to be done on the same computing cluster as is envisaged for the WBT $-$ it is unlikely that meaningful results can be obtained using one computer. Since this is related to the memory load, a single computer may well be unable to even allocate the required arrays.

Thus, individual portions of code should be copied and pasted into a separate algorithm to test that portion, while a combined test is likely to look very similar to an actual run. An important simplification in a test is that only a few MC trials need to be conducted, and perhaps just one would already help to fix various issues. Running a small number should clarify the cost.

\begin{table}
	\centering
	\caption{Estimated overall memory load on the master core, implications of which are discussed further in Section \ref{Overall_master_core_memory_load}.}
	\begin{tabular}{cc}
		\hline
		\multicolumn{2}{c}{Memory load on master core} \\
		\hline
		\multirow{2}{*}{Array} & Estimated size \\
		& (million entries) \\
		\hline
		Inner CB libraries & 197 \\
		Outer CB libraries & 249 \\
		Newtonian WB library & 57 \\
		MOND WB library & 263 \\
		Observed photographs & 5.4 \\
		\hline
		Total & 771 \\
		\hline
	\end{tabular}
	\label{Master_core_memory_load_summary}
\end{table}

\subsection{Interpolation of the CB libraries in $\gamma$}
\label{CB_gamma_interpolation}

Versions of the inner and outer CB libraries will be available for all 41 possible values of $\gamma$ covering the range $-2 \leq \gamma \leq 2$ in steps of 0.1. Once the MCMC algorithm has decided what $\gamma$ it wants to try on any particular iteration (Appendix \ref{Restricting_MCMC_range}), the CB libraries will be interpolated to this $\gamma$. We will use quadratic interpolation, so three library values of $\gamma$ must be considered. To maximize the accuracy, the central value will be the library value of $\gamma$ closest to the MCMC value, which will require a rounding operation. An exception will be made if the MCMC value of $\gamma$ is very close to either limit, to ensure the requested library values are available. For instance, if the MCMC sampler wishes to try $\gamma = -1.96$, then the used library values of $\gamma$ will be $-2$, $-1.9$, and $-1.8$; but for $\gamma = -1.76$, the used library values will be $-1.9$, $-1.8$, and $-1.7$. The accuracy should be somewhat better in the latter case as the library values are on average somewhat closer to the target, but this is difficult to avoid as $\gamma < -2$ is mathematically impossible (Equation \ref{gamma_definition}). Appropriate weights will be found for the three library values of $\gamma$.

\begin{table}
	\centering
	\caption{Estimated cost of marginalising the outer CB libraries over $\gamma$ (Section \ref{CB_gamma_interpolation}). This step will be parallelised over the underlined variable, allowing the inner CB libraries to be concurrently marginalised over $\gamma$. That step is not shown here because the inner CB libraries are smaller, so the cores used for them should finish earlier.}
	\begin{tabular}{cc}
		\hline
		\multicolumn{2}{c}{Outer CB library $\gamma$ interpolation cost} \\
		\hline
		Variable & Pixels \\
		\hline
		Libraries & 2 \\
		\underline{$\Delta \widetilde{M}$} & 15 \\
		Required values of $\gamma$ & 3 \\
		$\Delta \widetilde{v}_\parallel$ & 450 \\
		$\Delta \widetilde{v}_\perp$ & 450 \\
		Basic element (ns) & 50 \\
		Cores assumed & 15 \\
		\hline
		Estimated time (ms) & 61 \\
		\hline
	\end{tabular}
	\label{Outer_CB_library_gamma_interpolation_cost}
\end{table}

For simplicity, the parallelisation will only be over $\Delta \widetilde{M}$, so at most 15 cores can be used in parallel for the outer CB library. We will use another 15 in parallel for the inner CB library, but these cores should finish earlier as the inner CB library has only 400 pixels in $\Delta \widetilde{v}_{\parallel, \perp}$ rather than 450. The runtime will therefore be determined by the outer CB library, and is summarized in Table \ref{Outer_CB_library_gamma_interpolation_cost}. As the cost of this step is relatively modest, we do not envisage using a more complicated parallelisation scheme that employs more cores.

\subsubsection{Combining the CB libraries for the desired $f_{\text{CB}}$}
\label{Combine_CB_libraries}

Once the $\matr{P}_{\text{1CB}}$ and $\matr{P}_{\text{2CB}}$ libraries have been interpolated in $\gamma$, they must be combined to yield a single CB perturbation library
\begin{eqnarray}
    \matr{P}_{\text{CB}} ~=~ 2 f_{\text{CB}} \left( 1 - f_{\text{CB}} \right) \matr{P}_{\text{1CB}} \, + \, {f_{\text{CB}}}^2 \, \matr{P}_{\text{2CB}} \, .
\end{eqnarray}
The basic principle is to consider the possibility that neither, one, or both stars detected in a WB is/are contaminated by an undetected CB. To reduce the number of intermediate variables, we will jointly interpolate in $\gamma$ and combine the 1CB and 2CB libraries into a single CB library.

The probability of no CB contamination is $\left( 1 - f_{\text{CB}} \right)^2$. This case will be handled using a separate array that is added to the photo using a vastly simpler protocol which does not consider CB contamination, thereby skipping the complicated and relatively expensive stages described in Section \ref{Convolving_CB_WB_libraries}.

\subsection{Marginalising the WB libraries over $e$}
\label{Marginalise_WB_e}

As discussed in Section \ref{Stretch_factor_determination}, the WB $e$ is expected to change rapidly in MOND over only a few orbits \citep[see also figure 20 of][]{Banik_2018_Centauri}. While this is not the case in Newtonian gravity, the fact that it is true in one of the gravity theories being tested means that it is not very meaningful to consider a WB with fixed $e$. Instead, the WB orbit libraries should be marginalised over $e$ at an early stage in the evaluation of the cost function. This is not possible in advance of the MCMC analysis because the distribution over $e$ is one of the model parameters ($\gamma$ in Table \ref{Model_parameter_list}).

\subsubsection{Marginalising the Newtonian WB library}
\label{Marginalise_WB_Newton}

The Newtonian WB library does not have separate results for different $M$, leading to a relatively modest cost for marginalising it over $e$ (Table \ref{Marginalise_Newton_WB_cost}). It is therefore envisaged that parallelisation over only $\widetilde{r}_{\text{sky}}$ will be sufficient.

\begin{table}
	\centering
	\caption{Estimated cost of marginalising the Newtonian WB library over $e$ (Section \ref{Marginalise_WB_Newton}). This will be parallelised over the underlined variable.}
	\begin{tabular}{cc}
		\hline
		\multicolumn{2}{c}{Marginalising Newtonian WB library cost} \\
		\hline
		Variable & Pixels \\
		\hline
		$a$ & 39 \\
		$e$ & 100 \\
		\underline{$\widetilde{r}_{\text{sky}}$} & 100 \\
		$\widetilde{v}$ & 142 \\
		Basic element (ns) & 20 \\
		Cores assumed & 60 \\
		\hline
		Estimated time (ms) & 18 \\
		\hline
	\end{tabular}
	\label{Marginalise_Newton_WB_cost}
\end{table}

\subsubsection{Marginalising the MOND WB library}
\label{Marginalise_WB_MOND}

Table \ref{Marginalise_MOND_WB_cost} shows the estimated cost of marginalising the MOND WB library over $e$. As with the Newtonian case, the basic element is a vectorised dot product calculation, which is expected to have a rather low cost. This will almost certainly be dominated by reading from the shared WB library. For optimisation with the next stage (Section \ref{Gravity_law_interpolation}), this stage will also be parallelised over $\left(M, a \right)$.

\begin{table}
	\centering
	\caption{Estimated cost of marginalising the MOND WB library over $e$ (Section \ref{Marginalise_WB_MOND}). This will be parallelised over the underlined variables.}
	\begin{tabular}{cc}
		\hline
		\multicolumn{2}{c}{Marginalising MOND WB library cost} \\
		\hline
		Variable & Pixels \\
		\hline
		\underline{$M$} & 15 \\
		\underline{$a$} & 39 \\
		$e$ & 20 \\
		$\widetilde{r}_{\text{sky}}$ & 125 \\
		$\widetilde{v}$ & 180 \\
		Basic element (ns) & 20 \\
		Cores assumed & 60 \\
		\hline
		Estimated time (ms) & 88 \\
		\hline
	\end{tabular}
	\label{Marginalise_MOND_WB_cost}
\end{table}

\subsection{Applying stretch factors for the desired gravity law}
\label{Gravity_law_interpolation}

One of the most critical aspects of the WBT is obtaining an accurate WB orbit library for an arbitrary value of the gravity law parameter $\alpha_{\text{grav}}$ (Table \ref{Model_parameter_list}). The WB library is known in advance for the cases $\alpha_{\text{grav}} = 0$ and $\alpha_{\text{grav}} = 1$, corresponding to Newtonian and Milgromian gravity, respectively. These must serve as the starting point. For some general $\alpha_{\text{grav}}$, the appropriate stretch factor to the Newtonian probabilities is
\begin{eqnarray}
	\bm{S} ~=~ 1 + \alpha_{\text{grav}} \left( \bm{S}_{\text{MOND}} - 1 \right) \, ,
	\label{Stretch_factor_main}
\end{eqnarray}
where $\bm{S}_{\text{MOND}}$ is read in from the stretch factor library $\matr{S}$ (Section \ref{Stretch_factor_determination}) interpolated to the appropriate WB eccentricity distribution $\gamma$ using quadratic interpolation. Note that $\bm{S} \equiv \left( S_r, S_v \right)$ contains both the radial stretch factor $S_r$ and the velocity stretch factor $S_v$. A 2D stretch is required as the ranges in both $\widetilde{r}_{\text{sky}}$ and $\widetilde{v}_{\text{sky}}$ are typically $\approx 20\%$ higher in MOND than in Newtonian gravity. Equation \ref{Stretch_factor_main} recovers $\bm{S} = \bm{1}$ and $\bm{S} = \bm{S}_{\text{MOND}}$ when $\alpha_{\text{grav}} = 0$ and $\alpha_{\text{grav}} = 1$, respectively.

To reduce the already high memory load on the master core (Table \ref{Master_core_memory_load_summary}), the residuals between the stretched Newtonian array and the MOND array will be calculated during the MCMC stage. This yields a corrections array $\matr{C}$ that would give the MOND array if added to the Newtonian array stretched by $\bm{S}_{\text{MOND}}$. Note that $\matr{C}$ necessarily sums to zero, making its normalisation more difficult to check.

For some general $\alpha_{\text{grav}}$, the corrections should themselves be stretched to avoid creating contributions at e.g. $\widetilde{v} = 1.6$ when $\alpha_{\text{grav}} = 0.01$. The corrections will also be scaled by $\alpha_{\text{grav}}$ so that no further corrections are applied in the Newtonian case. The appropriate stretch factor for the corrections is
\begin{eqnarray}
	\bm{S}_c ~=~ {\bm{S}_{\text{MOND}}}^{-1} + \alpha_{\text{grav}} \left( 1 - {\bm{S}_{\text{MOND}}}^{-1} \right) \, .
	\label{Stretch_factor_correction}
\end{eqnarray}
This ensures the corrections are stretched by ${\bm{S}_{\text{MOND}}}^{-1}$ in the Newtonian case and 1 in the MOND case, with ${\bm{S}_{\text{MOND}}}^{-1}$ denoting an element-wise inverse.

Combining these results, the distribution of $\left( \widetilde{r}_{\text{sky}}, \widetilde{v} \right)$ is
\begin{eqnarray}
	&& \matr{P} ~=~ \matr{P}_N \left( \overleftrightarrow{1 + \alpha_{\text{grav}} \left( \bm{S}_{\text{MOND}} - 1 \right)} \right) \\
	&+& \alpha_{\text{grav}} \matr{C} \left( \overleftrightarrow{{\bm{S}_{\text{MOND}}}^{-1} + \alpha_{\text{grav}} \left( 1 - {\bm{S}_{\text{MOND}}}^{-1} \right)} \right) \, , \nonumber
\end{eqnarray}
where $\overleftrightarrow{\bm{S}}$ denotes a 2D stretch by a factor $\bm{S}$. After $\matr{P}$ has been calculated, negative entries will be set to zero and the array renormalised.

Since it is important to include the full $\widetilde{v}$ range permissible in the gravity law under consideration, there will be no leakage of probability outside some `allowed' range of $\widetilde{v}$. This will be ensured with an appropriately wide choice of the $\widetilde{v}$ range considered by the algorithm. The only limit is that it will not reliably handle the situation in case $\widetilde{v} > 5$, but this is not expected to occur for realistic values of $\alpha_{\text{grav}}$ prior to including contamination from the CB (Section \ref{Convolving_CB_WB_libraries}).

Based on the information from Section \ref{Max_indices_WB} on the maximum $\widetilde{r}_{\text{sky}}$ and $\widetilde{v}$ that arise in the Newtonian and MOND WB libraries for each $\left(M, a \right)$, additional savings will be attempted by restricting the range of indices over which evaluations are done. These index ranges will themselves need to be scaled by the appropriate $\bm{S}$.

To summarize, three stretches are required:
\begin{enumerate}
	\item the Newtonian WB library by $\bm{S}_{\text{MOND}}$, the optimal factor to mimic the MOND distribution (found in Section \ref{Stretch_factor_determination} and quadratically interpolated to the desired $\gamma$);
	\item the Newtonian WB library by $\bm{S}$ (Equation \ref{Stretch_factor_main}); and
	\item the corrections array by $\bm{S}_c$ (Equation \ref{Stretch_factor_correction}).
\end{enumerate}
The first of these can in principle be avoided by calculating the corrections array in advance of the MCMC analysis, but this is not envisaged because it would significantly raise the already high memory load on the master core (Section \ref{Master_core_memory_load}).

These 2D stretches can be implemented by applying the appropriate $S_r$ and $S_v$ to each pixel, but this would entail significant overheads associated with finding the appropriate output pixels and their overlap fractions using relatively expensive rounding operations (Table \ref{Basic_operations_cost}). Therefore, it is envisaged that each stretch will be implemented in two stages, first along $\widetilde{r}_{\text{sky}}$ and then along $\widetilde{v}$. At each stage, a whole row or column will be stretched, spreading the cost of the above-mentioned overheads across a large number of elements. Essentially, the determination of the output pixels and their overlap fractions will be done only along the left and bottom edges of the array being stretched, rather than for every element. Further modest savings may accrue from vectorisation.

\begin{table}
	\centering
	\caption{Estimated cost of applying the stretch factors to the original WB libraries to obtain one suitable for the desired $\alpha_{\text{grav}}$ (Section \ref{Gravity_law_interpolation}). This step will be parallelised over the underlined variables. The resulting WB library will be sent to a shared array holding results for all ($M, a$), which can then be marginalised over $M$ (Section \ref{Marginalising_WB_over_M}). In both cases, communications with the shared array do not need special protection through the $critical$ directive, likely smoothing the operation of the algorithm. Although only 3 stretches are required, 4 are assumed for the cost estimate due to the higher cost of stretching the corrections array, which arises from the larger ($\widetilde{r}_{\text{sky}}$, $widetilde{v}$) range in MOND.}
	\begin{tabular}{cc}
		\hline
		\multicolumn{2}{c}{Stretch factor application cost} \\
		\hline
		Variable & Pixels \\
		\hline
		\underline{$M$} & 15 \\
		\underline{$a$} & 39 \\
		$\widetilde{r}_{\text{sky}}$ & 100 \\
		$\widetilde{v}$ & 142 \\
		Output pixels & 5 (average) \\
		Stretches & 4 ($\times$ Newtonian cost) \\
		Basic element (ns) & 50 \\
		Cores assumed & 60 \\
		\hline
		Estimated time (ms) & 138 \\
		\hline
	\end{tabular}
	\label{Stretch_factor_application_cost}
\end{table}

Table \ref{Stretch_factor_application_cost} summarizes the cost of obtaining the WB library for the desired $\alpha_{\text{grav}}$. Though only 3 stretches are required, we have assumed 4 because the MOND distribution $-$ and thus the corrections array $-$ would be wider than the Newtonian distribution, whose size is used in the cost estimate. The basic element is mostly a vectorised read and write, since the input vector can be stored in a temporary vector that is scaled to the appropriate overlap fraction. Given also that the arrays being incremented would be fairly small and thus private to each core, the basic element is assigned a rather low cost. For a stretch factor slightly exceeding 1, each input pixel will be spread over 2 or perhaps 3 pixels at each stretching stage, so 5 output pixels are assumed once the 2D nature of each stretch is considered. A small amount of time will also be required to read in the Newtonian WB library marginalised over $e$ for the particular $a$ under consideration, thereby yielding a private copy of this portion of the shared array. This private array is heavily used in the steps described in this section, but should be relatively cheap to set up in comparison to the other costs. Since the MOND WB library will not be stretched, it does not need to be read into a private array.

\subsection{Marginalising stretched WB libraries over $M$}
\label{Marginalising_WB_over_M}

Once the WB library has been determined for the desired $\alpha_{\text{grav}}$, it will be marginalised over $M$ using the distribution of WB total detected system mass $M$, which should be clear quite early in the project. At this stage, the resolution in velocity will also be reduced $4\times$ as a resolution of 0.01 is vastly better than in the photo (Table \ref{Pixel_decision_v}), and thus unnecessary. Since the original WB libraries will already have been marginalised over $e$ at an earlier stage (Section \ref{Marginalise_WB_e}), there will now be one less rank to these tensors, so the cost of marginalising over $M$ is negligible in comparison to the other costs (Table \ref{Marginalise_WB_over_M_cost}). The result will be a distribution over $\left( a, \widetilde{r}_{\text{sky}}, \widetilde{v}_{\text{sky}} \right)$, which must be stored in a shared array that different cores write to. Because of the low cost and possible parallelisation overheads, it is envisaged that this step will be parallelised only over $a$, which also removes the need to use the $critical$ directive.

\begin{table}
	\centering
	\caption{Estimated cost of marginalising the WB library over $M$ (Section \ref{Marginalising_WB_over_M}), parallelised over the underlined variable. The basic element consists of a vectorised dot product operation requiring a read from a shared array.}
	\begin{tabular}{cc}
		\hline
		\multicolumn{2}{c}{Marginalising over $M$ cost} \\
		\hline
		Variable & Pixels \\
		\hline
		$M$ & 15 \\
		\underline{$a$} & 39 \\
		$\widetilde{r}_{\text{sky}}$ & 150 \\
		$\widetilde{v}$ & 200 \\
		Basic element (ns) & 50 \\
		Cores assumed & 39 \\
		\hline
		Estimated time (ms) & 23 \\
		\hline
	\end{tabular}
	\label{Marginalise_WB_over_M_cost}
\end{table}

\subsection{Bunching WB libraries in $a$}
\label{a_bunching}

The majority of the computational cost of the cost function arises from calculations parallelised over the WB $\widetilde{r}_{\text{sky}}$. Therefore, the WB probability distribution will first be read in for each $\widetilde{r}_{\text{sky}}$ into a private 2D array covering $\left( a, \widetilde{v}_{\text{sky}} \right)$. For fixed $\widetilde{r}_{\text{sky}}$, the $r_{\text{sky}}$ bin in the photo is determined solely by $a$. Since the photo has rather wide bins in $r_{\text{sky}}$ (Table \ref{Pixel_decision_r}) and the subsequent steps are similar for all $a$ (Sections \ref{Scaling_CB_contamination_pattern_for_rtsky} and \ref{Convolving_CB_WB_libraries}), many different values of $a$ can be bunched together.

To maximize the accuracy, we will consider a forward binning protocol (Equation \ref{Forwards_binning}) which considers the full range in $a$ over each pixel. This is because the steps in $a$ are generally much wider than in $\widetilde{r}_{\text{sky}}$, so it is less helpful to consider the full range of $\widetilde{r}_{\text{sky}}$ across the pixel of interest. Instead, the central value of $\widetilde{r}_{\text{sky}}$ will be assumed. To handle the finite range in $a$, the basic idea is that if the input pixel covers say $a = 11-12$ kAU and $\widetilde{r}_{\text{sky}} = 1.2$, then the pixel should be smeared out over the range $13.2-14.4$ kAU. Assuming a uniform distribution over the input pixel and output bins covering each kAU in $r_{\text{sky}}$, this means $0.8/1.2$ of the probability should be assigned to the output pixel covering $r_{\text{sky}} = 13-14$ kAU as the overlap range is 0.8 kAU and the full width after stretching is 1.2 kAU. In this case, the remainder of the probability should be assigned to the next output pixel in $r_{\text{sky}}$, though in general it is possible that there will be further output pixels to consider. We assume that each $a$ contributes to 2.5 output pixels in $r_{\text{sky}}$ on average. The result will be that instead of having a WB library over $\left( a, \widetilde{v} \right)$, we will have a library over $\left( r_{\text{sky}}, \widetilde{v} \right)$, which greatly simplifies the process of building up the simulated photo as the bins in $r_{\text{sky}}$ correspond to those in the photo.

Since it is quite possible that not all $r_{\text{sky}}$ bins receive a contribution even when all $a$ values are considered, we will set up a vector of logicals containing information on whether each photographic $r_{\text{sky}}$ bin receives a contribution from at least one $a$. This array of logicals must be reset when considering a new value of $\widetilde{r}_{\text{sky}}$. It will be used to skip updates to those photographic $r_{\text{sky}}$ bins (Table \ref{Pixel_decision_r}) which do not receive any contribution, for example very high $r_{\text{sky}}$ bins in a case with very low $\widetilde{r}_{\text{sky}}$. We assume that on average, only 15 of the 18 photographic $r_{\text{sky}}$ bins receive a non-zero contribution.

\begin{table}
	\centering
	\caption{Estimated cost of bunching the WB library in $a$ for each $\widetilde{r}_{\text{sky}}$ according to the output radial bin in the photo (Section \ref{a_bunching}). This step will be parallelised over the underlined variable, which should be sufficient given the low estimated cost.}
	\begin{tabular}{cc}
		\hline
		\multicolumn{2}{c}{Bunching WB library in $a$ cost} \\
		\hline
		Variable & Pixels \\
		\hline
		\underline{$\widetilde{r}_{\text{sky}}$} & 120 (average) \\
		$a$ & 39 \\
		Output pixels & 2.5 (average) \\
		$\widetilde{v}$ & 70 (average) \\
		Basic element (ns) & 30 \\
		Cores assumed & 60 \\
		\hline
		Estimated time (ms) & 0.4 \\
		\hline
	\end{tabular}
	\label{Bunching_WB_library_in_a_cost}
\end{table}

Table \ref{Bunching_WB_library_in_a_cost} summarizes the computational cost of bunching the WB library in $a$. The very low cost justifies the use of additional measures to improve the accuracy. Although this stage is relatively cheap, it is still crucial to the efficient operation of the relatively more expensive stage described in Section \ref{Velocity_transformation_application} because instead of considering all 39 pixels in $a$, on average only 15 pixels in $r_{\text{sky}}$ need to be considered. To optimise this stage, we will use a private array to hold the `raw' WB library for each $\widetilde{r}_{\text{sky}}$ after bunching in $a$.

\subsection{Scaling the CB contamination pattern to the desired $\widetilde{r}_{\text{sky}}$ and $C_{\text{limit}}$}
\label{Scaling_CB_contamination_pattern_for_rtsky}

The $\matr{P}_{\text{CB}}$ library found in Section \ref{CB_gamma_interpolation} is hypothecated to the case that $a_{\text{int, max}}$ for the CB is equal to the WB $r_{\text{sky}}$. Our model does not work in this way, so the $\matr{P}_{\text{CB}}$ library must be stretched in velocity. The appropriate factor is set by $\widetilde{r}_{\text{sky}}$ and the chaos limit parameter $C_{\text{limit}}$, defined so
\begin{eqnarray}
    a_{\text{int, max}} ~\equiv~ a C_{\text{limit}} \, ,
\end{eqnarray}
where $a$ refers to the WB. A good assumption is that the CB becomes chaotic if $a_{\text{int}} \ga a/3$, so $C_{\text{limit}} \approx 1/3$. This will be treated as a free parameter in the MCMC analysis (Table \ref{Model_parameter_list}).

\subsubsection{Backward binning}

The CB velocity perturbations should be scaled down for a high $C_{\text{limit}}$ and up for a high $\widetilde{r}_{\text{sky}}$. This is because wider CBs have a lower orbital velocity, while the expected WB circular velocity is reduced at high $r_{\text{sky}}$, increasing the impact of the CB on the observed WB $\widetilde{v}$. As a result, the CB library should be stretched in velocity by
\begin{eqnarray}
    f ~=~ \sqrt{\frac{\widetilde{r}_{\text{sky}}}{C_{\text{limit}}}} \, .
    \label{CB_stretch_factor}
\end{eqnarray}

Since there are many pixels with $\widetilde{r}_{\text{sky}} \ll 1$, it becomes necessary to consider a very wide range of velocity in the CB library to build up the CB contamination pattern over the desired range. While this is possible in principle because the CB library covers out to $\Delta \widetilde{v}_{\perp, \parallel}$ of 90, the computational cost of a standard forward binning protocol would be prohibitively high, especially as each input pixel would generally contribute to several output pixels. However, the probability density should change little between e.g. $\Delta \widetilde{v}_\parallel$ of 87.2 and 87.8 when $\Delta \widetilde{v}_\perp = 70$. This is not exploited in the usual approach where the corner $\left(x, y \right)$ arguments in some bin of the input matrix $\matr{I}$ are scaled by the relevant factor $f$, the results are rounded to find the bins in the results matrix $\matr{R}$, and these bins are incremented appropriately. This procedure can be summarized as
\begin{eqnarray}
	\matr{R} \left( \overbrace{fx_i}^{x_r}, \overbrace{fy_i}^{y_r} \right) ~+=~ \matr{I} \left( x_i, y_i \right) \, ,
	\label{Forwards_binning}
\end{eqnarray}
with the subscripts $i$ or $r$ denoting input or results, respectively.

To reduce the computational cost by reducing the accuracy somewhat at very high $\Delta \widetilde{v}$ in the CB library, we will use an alternative protocol where instead of calculating forward from some input matrix $\matr{I}$ to the results matrix $\matr{R}$, the probability in each pixel of $\matr{R}$ is found by looking up which pixels in $\matr{I}$ contribute to it. The basic principle is that if the input matrix is scaled by some factor $f$, the contribution to some pixel $\matr{R} \left( x_r, y_r \right)$ arises from $\matr{I} \left( \frac{x_r}{f}, \frac{y_r}{f} \right)$. In general, there is no pixel at this location, so interpolation is required to estimate the probability density there. Importantly, the probability in $\matr{I}$ must also be scaled by the inverse of the Jacobian determinant, so
\begin{eqnarray}
	\matr{R} \left( x_r, y_r \right) ~=~ \frac{\matr{I} \left( \frac{x_r}{f}, \frac{y_r}{f} \right)}{f^2} \, .
	\label{Backwards_binning}
\end{eqnarray}

The main goal is to improve the efficiency when $f < 1$. If instead $f > 1$, this `backward binning' procedure will actually be more accurate than the standard forward protocol (Equation \ref{Forwards_binning}) as small-scale structure in the input library is preserved by linear interpolation. This is helpful because for the WBT, it is critical to consider WBs with high $\widetilde{v}$, which requires a high $\widetilde{r}_{\text{sky}}$. Since $C_{\text{limit}} < 1$, such systems have $f \ga 1$ (Equation \ref{CB_stretch_factor}).

\subsubsection{Nearest corner interpolation}

To maximize the accuracy of the interpolation, we will use what we term `nearest corner interpolation'. The idea is that if values are available on every integer, estimating the value at $\left( 0.1, 0.1 \right)$ uses the evaluations at $\left( 0, 0 \right)$, $\left( 0, 1 \right)$, and $\left( 1, 0 \right)$. However, estimating the value at $\left( 0.1, 0.9 \right)$ uses the evaluations at $\left( 0, 1 \right)$, $\left( 0, 0 \right)$, and $\left( 1, 1 \right)$, with the baseline pixel given first in both cases. Compared to interpolating towards the upper right in all cases, this should typically halve the distance between the point where the value is required and the baseline pixel, which now becomes the nearest point where the value is known. The difference in value relative to the baseline determination is estimated using gradients obtained by finite differencing. Halving the distance to the baseline should reduce the error $4\times$, which alternatively allows a lower resolution to be used for similarly accurate results. This technique already underpins the force evaluations required to prepare the MOND WB library (Section \ref{Directly_relevant_prior_works}).

\subsubsection{Finite area correction}

A further significant improvement to the accuracy is possible especially in the case $f < 1$, when it could be inaccurate to approximate the average probability density over the large input region by the value at its centre. The basic principle is to also use the values at the four corners of this region in order to include second order corrections. We have found analytically and numerically that the optimal strategy is to weight the central determination of the probability density at $8\times$ the corner determinations, so the weight of the central determination is $2/3$ while that of all four corner determinations combined is $1/3$. This is equivalent to approximating that
\begin{eqnarray}
    &&\frac{\int_{x_0 - dx}^{x_0 + dx} \int_{y_0 - dy}^{y_0 + dy} f \left( x, y \right) \, dx \, dy}{4 \, dx \, dy} ~=~ \frac{1}{12} \left[ 8 f \left( x_0, y_0 \right) \right. \nonumber \\
    &&+ f \left( x_0 - dx, y_0 - dy \right) + f \left( x_0 - dx, y_0 + dy \right) \nonumber \\
    &&+ \left. f \left( x_0 + dx, y_0 - dy \right) + f \left( x_0 + dx, y_0 + dy \right) \right] \, .
    \label{Accurate_2D_integral}
\end{eqnarray}

\begin{table}
	\centering
	\caption{Estimated cost of determining the CB contamination pattern (Section \ref{Scaling_CB_contamination_pattern_for_rtsky}), parallelised over the underlined variable. Despite the relatively high cost, significant inefficiencies would ensue from further parallelisation, in particular over $\Delta \widetilde{M}$.}
	\begin{tabular}{cc}
		\hline
		\multicolumn{2}{c}{Determining CB contamination pattern cost} \\
		\hline
		Variable & Pixels \\
		\hline
		\underline{$\widetilde{r}_{\text{sky}}$} & 120 \\ [3pt]
		$\Delta \widetilde{M}$ & 15 \\
		$\Delta \widetilde{v}_\perp$ & 125 \\
		$\Delta \widetilde{v}_\parallel$ & 200 \\
		Basic element (ns) & $100+150$ \\
		Cores assumed & 60 \\
		\hline
		Estimated time (ms) & 188 \\
		\hline
	\end{tabular}
	\label{CB_contamination_pattern_determination_cost}
\end{table}

To avoid repeated corner determinations, we will first find the probability density in the CB library at the corner locations for each desired pixel in the CB contamination pattern. This will require loops where the range of each velocity index is 1 more than for the CB contamination pattern. We will then obtain the central determination, and combine it with the previously recorded corner determinations using the above-mentioned relative weights. This two-step procedure is estimated to have a basic element cost of 100 ns in the first step and 150 ns in the second step, leading to the determination of the CB contamination pattern having an estimated cost summarized in Table \ref{CB_contamination_pattern_determination_cost}. The high estimated basic element cost is due to reading in from the large shared $\matr{P}_{\text{CB}}$ array, though the corner values will be stored in a private array that should be much faster to read from. It is envisaged that all $\Delta \widetilde{M}$ will be considered jointly, so the costs associated with finding the appropriate indices in the input array should be negligible once shared over all 15 pixels in $\Delta \widetilde{M}$.

\subsection{Convolving the CB and WB libraries}
\label{Convolving_CB_WB_libraries}

One of the most expensive stages of the cost function is sure to be the addition of the CB velocity perturbation to the WB $\widetilde{\bm{v}}$, since this requires loops over parameters of both the CB and the WB, which are essentially independent. The feasibility of the whole plan relies fundamentally on whether this step is affordable, which in turn relies on keeping the number of loops to an absolute minimum. At least 5 variables must be considered at this stage: 2 from the WB $\left( r_{\text{sky}}, \widetilde{v} \right)$ required to know which part of the simulated photo to update, and 3 from the CB $\left( \Delta \widetilde{M}, \Delta \widetilde{v}_\perp, \Delta \widetilde{v}_\parallel \right)$ which hold in the most compressed form the impact of CBs on observables. The convolution of the CB and WB libraries will be split into the stages described below.

\subsubsection{Stretching the WB $\widetilde{v}$ for hidden mass}
\label{Hidden_mass_stretching}

To account for the hidden mass effect of CB(s), it is necessary to scale the WB $\widetilde{v}$ up by a factor of $\sqrt{1 + \Delta \widetilde{M}}$ (Equation \ref{Hidden_mass_effect}). This will be implemented using a standard 1D stretch in $\Delta \widetilde{M}$, though considering the full width of the input $\widetilde{v}$ pixel to maximize the accuracy. For simplicity, we will only consider the bin centre value of $\Delta \widetilde{M}$. Since the resolution in $\widetilde{v}$ of 0.04 is very good relative to the pixels in the photo (Table \ref{Pixel_decision_v}), this should be sufficiently accurate.

For a stretch factor $>1$, each input $\widetilde{v}$ pixel must yield at least 2 output pixels and possibly 3, though not more as the stretch factor will be $<1.3$ for a maximum pixel centre $\Delta \widetilde{M}$ of 0.675. Therefore, we assume 2.5 output pixels on average. This leads to the low estimated cost summarized in Table \ref{Hidden_mass_stretching_cost}, which justifies the use of additional measures to improve the accuracy.

\begin{table}
	\centering
	\caption{Estimated cost of stretching the `raw' WB library for hidden mass in the CB (Section \ref{Hidden_mass_stretching}), parallelised over the underlined variable.}
	\begin{tabular}{cc}
		\hline
		\multicolumn{2}{c}{Stretching for hidden mass cost} \\
		\hline
		Variable & Pixels \\
		\hline
		\underline{$\widetilde{r}_{\text{sky}}$} & 120 (average) \\ [3pt]
		$\Delta \widetilde{M}$ & 15 \\
		$r_{\text{sky}}$ & 15 (average, 18 in photo) \\
		Input $\widetilde{v}$ & 70 \\
		Output pixels & 2.5 (average) \\
		Basic element (ns) & 30 \\
		Cores assumed & 60 \\
		\hline
		Estimated time (ms) & 2.4 \\
		\hline
	\end{tabular}
	\label{Hidden_mass_stretching_cost}
\end{table}

\subsubsection{Determining the velocity transformation between WB and total $\widetilde{v}$}
\label{Velocity_transformation_determination}

In addition to inflating the WB mass beyond the observational estimate, the other major effect of the CB is the induced recoil velocity on the contaminated star, or more precisely the induced photocentre-barycentre offset due to the non-linear mass-luminosity relation \citep[$L \appropto M^{5.5}$ in the Gaia band, equation 4.3 of][]{Pittordis_thesis}. The recoil velocity (Equation \ref{q_scaling_v}) must be added to the WB relative velocity using a standard Galilean transformation. Since the important quantity in the end is the total $\widetilde{v}$, we need to distinguish whether the induced recoil velocity is parallel or orthogonal to the WB $\widetilde{\bm{v}}_{\text{sky}}$.

An important simplification is that the manner in which the CB affects the problem is independent of the WB $a$, since this dependence is cancelled out by the use of $\widetilde{r}_{\text{sky}}$ rather than $r_{\text{sky}}/a$ in Equation \ref{CB_stretch_factor}. However, the physics of the WB does depend on $a$ due to the adoption of a 50 AU floor and 100 kAU ceiling on the WB separation during the orbit integration \citep{Banik_2018_Centauri}. Therefore, the relatively expensive procedure of convolving the CB and WB libraries is very similar for all $a$, but must nonetheless be applied to WBs with different $a$.

To exploit this, we will set up a transformation matrix which will hold the probability that some input pixel in $\widetilde{v}$ contributes to some output $\widetilde{v}$ pixel in the photo. The entries for fixed input $\widetilde{v}$ need not sum to 1 because the CB might induce a very large perturbation that would cause observers to reject the WB. However, this is not likely as the photo covers out to $\widetilde{v} = 5$ (Table \ref{Pixel_decision_v}), so very little of the probability in each input $\widetilde{v}$ pixel should be lost, especially at low $\widetilde{r}_{\text{sky}}$.

The total $\widetilde{v}$ corresponding to any given total $\left( \widetilde{v}_\perp, \widetilde{v}_\parallel \right)$ pair must be found using the Pythagoras Theorem. To improve the efficiency, this step will be done in advance of the MCMC stage, when we will also determine whether the pixel centre $\widetilde{v} > 5$ and thus does not contribute to the simulated photo. It then becomes necessary to work in the space of total (WB $+$ CB) $\widetilde{\bm{v}}$, though we will not explicitly determine the probability distribution in this space as this depends on the WB distribution of $\widetilde{v}$, defeating the purpose of creating a velocity transformation. Each pixel in this space corresponds to a known velocity index in the photo, and thus in the velocity transformation. The index for the input $\widetilde{v}$ depends on how much of the total $\widetilde{v}_\parallel$ is contributed by the WB. As a result, dividing the workload amongst pixels in the 2D space of total $\widetilde{\bm{v}}$ means that each time we will be considering contributions to only one column in the transformation matrix, corresponding to a fixed output velocity index. As we consider larger values of $\widetilde{v}_\parallel$ in the WB, we simultaneously need to consider smaller values of $\Delta \widetilde{v}_\parallel$ from the CB such that the total remains the same. Therefore, a vector drawn from the CB contamination pattern at fixed $\Delta \widetilde{v}_\perp$ contributes directly to a vector in the transformation matrix at fixed output $\widetilde{v}$, albeit in reverse order. Note that the likelihood of different WB $\widetilde{v}$ does not enter here as the idea is to set up a velocity transformation matrix. The only relevant aspect of the actual WB $\widetilde{v}$ distribution is the index range for the highest $\Delta \widetilde{M}$, as this sets the required input $\widetilde{v}$ range of the velocity transformation.

A major efficiency saving is possible because the velocity pixels in the photo are quite wide due to the finite number of observed WBs (Table \ref{Pixel_decision_v}), but the calculations until now have used rather narrow bins in $\widetilde{v}$ to maintain good accuracy. As a result, many adjacent pixels in total $\widetilde{v}_\parallel$ would contribute to the same velocity index in the photo, and thus to the same column in the velocity transformation. This allows us to consider e.g. 10 adjacent pixels with fixed total $\widetilde{v}_\perp$ but different $\widetilde{v}_\parallel$. The contribution to the velocity transformation for the lowest input $\widetilde{v}$ index is the sum of the above-mentioned 10 pixels in the CB contamination pattern, which e.g. have $\Delta \widetilde{v}_\parallel$ indices in the range $101-110$. For the next input $\widetilde{v}$ index in the velocity transformation but still at the same output $\widetilde{v}$, the contribution is the sum of entries in the CB contamination pattern with the same $\Delta \widetilde{v}_\perp$ as before but with $\Delta \widetilde{v}_\parallel$ indices in the range $100-109$. Since the sum over indices $100-109$ only differs slightly from the sum over indices $101-110$, it is not necessary to simply sum 10 numbers in both cases. This is required for the first case ($101-110$), but if the sum is stored as a scalar, then the result for the second case ($100-109$) can instead be found by subtracting the value at index 110 and adding the value at index 100. This will require only 2 addition/subtraction operations rather than 10. The savings are likely to be significant because the transformation matrix will probably need to be rated for $\approx 70$ different input $\widetilde{v}$ indices in order to cover the range for which it will be needed, and this procedure should significantly reduce the cost in 69 of these cases.

With this major efficiency saving, the computational cost of finding the velocity transformation is estimated in Table \ref{Velocity_transformation_determination_cost}. The standard approach would be to consider 225 pixels in total $\widetilde{v}_\parallel$ and 125 pixels in $\Delta \widetilde{v}_\perp$. Even with the geometric efficiency saving of $\mathrm{\pi}/4$, this would still involve considering $\approx 22,000$ pixels. However, we only need to consider 2275 `line operations' if we jointly consider adjacent pixels in total $\widetilde{v}_\parallel$ as far as possible while considering different $\widetilde{v}_\perp$ separately. Apart from edge cases, the basic element thus consists of two reads and two addition/subtraction operations to find the appropriate increment to the velocity transformation starting from the previous increment, with a further read, $+$, and write required on the velocity transformation to actually update it. The basic element is assumed to have a relatively low cost because the arrays being read from and written to should be quite small, and are therefore expected to be private.

\begin{table}
	\centering
	\caption{Estimated cost of determining the velocity transformation relating the WB and total $\widetilde{v}$ (Section \ref{Velocity_transformation_determination}), parallelised over the underlined variable. The basic principle is to consider the space of total $\left( \widetilde{v}_\perp, \widetilde{v}_\parallel \right)$. We do so by bunching adjacent pixels in $\widetilde{v}_\parallel$ at fixed $\widetilde{v}_\perp$ that map to the same velocity index in the photo. These `line operations' drastically reduce the estimated cost.}
	\begin{tabular}{cc}
		\hline
		\multicolumn{2}{c}{Velocity transformation determination cost} \\
		\hline
		Variable & Pixels \\
		\hline
		\underline{$\widetilde{r}_{\text{sky}}$} & 120 (average) \\ [3pt]
		$\Delta \widetilde{M}$ & 15 \\
		Line operations $\left( \widetilde{v}_\perp, \widetilde{v}_\parallel \right)$ & 2275 \\
		Input $\widetilde{v}$ & 70 \\
		Basic element (ns) & 35 \\
		Cores assumed & 60 \\
		\hline
		Estimated time (ms) & 167 \\
		\hline
	\end{tabular}
	\label{Velocity_transformation_determination_cost}
\end{table}

Each $\widetilde{v}$ bin in the photo corresponds to an annulus in the space of total $\left( \widetilde{v}_\perp, \widetilde{v}_\parallel \right)$. Therefore, it is not completely accurate to assign a pixel in total $\left( \widetilde{v}_\perp, \widetilde{v}_\parallel \right)$ to a single $\widetilde{v}$ bin in the photo. This is approximately corrected using the procedure described in Appendix \ref{Pixel_area_corrections}.

Another inaccuracy is that summing a pixel where e.g. the WB has $\widetilde{v}$ in the range $2-3$ with a pixel where the CB has $\Delta \widetilde{v}_\parallel$ in the range $10-11$ is assumed to yield a total $\widetilde{v}_\parallel$ entirely within the pixel $12-13$. In reality, supposing a uniform distribution over the input pixels, half the probability should be assigned to the output pixel with $\widetilde{v}_\parallel = 12-13$, and half to the next pixel covering $13-14$. This is corrected on the input side as described in Section \ref{Velocity_transformation_application}.

The velocity transformation between WB and total $\widetilde{v}$ does not apply to scenarios like the 0CB case where $\Delta \bm{\widetilde{v}} = \bm{0}$, avoiding the need for a reduction from a 2D array to a 1D array using annular bins. We will therefore use a separate array labelled `DI' (for direct injection) holding contributions to the partial photo which should not be corrected using the procedure described in Appendix \ref{Pixel_area_corrections}, but should instead be directly injected into the simulated photo. Though the major contribution to this is from the 0CB case, it also applies to cases with CB contamination if all involved CBs have an exactly equal mass ratio, leading to no photocentre-barycentre offset assuming an unresolved CB (a resolved CB should lead to prompt rejection by observers). The protocol for these cases is described further in Appendix \ref{Equal_mass_CBs}, and relies critically on the DI array to circumvent steps associated with the annular binning that is otherwise required.

\subsubsection{Applying the velocity transformation}
\label{Velocity_transformation_application}

Since the WB has $\widetilde{v}_\perp = 0$ by construction, summing the CB and WB orbital velocities in this direction is trivial $-$ only the CB contributes. In the $\widetilde{v}_\parallel$ direction, we need to handle the issue raised in Section \ref{Velocity_transformation_determination} with combining the CB and WB contributions. Since the velocity transformation is already fairly expensive (Table \ref{Velocity_transformation_determination_cost}), this issue should not be fixed by altering the transformation. Instead, the fix will be in the input array of $\widetilde{v}$ from the WB, possibly stretched for hidden mass as described in Section \ref{Hidden_mass_stretching}. \textbf{The probability in each input $\widetilde{v}$ pixel will be divided into equal contributions to the original pixel and the next pixel.} This has the effect of extending by one the range of non-zero pixels in $\widetilde{v}$ on the input side. The computational cost of this dithering protocol is assumed to be negligible in comparison to the other costs.

It should be quite accurate to apply the velocity transformation (Section \ref{Velocity_transformation_determination}) to the dithered input $\widetilde{v}$ distribution. We will reduce the cost by first combining all contributions with fixed output $\widetilde{v}$ index in the photo, before incrementing that pixel in the partial photo for each $\widetilde{r}_{\text{sky}}$ (the partial photos will be combined into the simulated photo as described in Section \ref{Updating_sim_photo}). At fixed output $\widetilde{v}$, the velocity transformation holds the likelihood that different input $\widetilde{v}$ contribute to this particular velocity index in the photo. The actual contribution is then just the dot product between this list of likelihoods and the probability distribution over the input $\widetilde{v}$. Bearing in mind that the evaluation of this dot product should be the main cost of this stage, the estimated cost of applying the velocity transformation is shown in Table \ref{Velocity_transformation_application_cost}.

The DI array described in Section \ref{Velocity_transformation_determination} will also need to be injected into the simulated photo, but the cost of this is assumed to be negligible in comparison to the other costs because each input velocity pixel maps to only one velocity pixel in the photo. Moreover, the DI array will only be used to increment the photo on the last $\Delta \widetilde{M}$ index (Appendix \ref{Equal_mass_CBs_MCMC_adjustments}).

\begin{table}
	\centering
	\caption{The estimated cost of applying the velocity transformation between WB and total $\widetilde{v}$ (Section \ref{Velocity_transformation_application}), parallelised over the underlined variable. We have assumed that not all $r_{\text{sky}}$ indices in the photo receive a contribution for every $\widetilde{r}_{\text{sky}}$.}
	\begin{tabular}{cc}
		\hline
		\multicolumn{2}{c}{Velocity transformation application cost} \\
		\hline
		Variable & Pixels \\
		\hline
		\underline{$\widetilde{r}_{\text{sky}}$} & 120 (average) \\ [3pt]
		$\Delta \widetilde{M}$ & 15 \\
		$r_{\text{sky}}$ & 15 (average, 18 in photo) \\
		Input $\widetilde{v}$ & 70 \\
		Output $\widetilde{v}$ & 30 (same as photo) \\
		Basic element (ns) & 50 \\
		Cores assumed & 60 \\
		\hline
		Estimated time (ms) & 47 \\
		\hline
	\end{tabular}
	\label{Velocity_transformation_application_cost}
\end{table}

\subsection{Updating the simulated photograph}
\label{Updating_sim_photo}

Since many of the calculations described so far will be parallelised over $\widetilde{r}_{\text{sky}}$, each core will have a private partial photograph holding contributions to the simulated photo arising from only the particular $\widetilde{r}_{\text{sky}}$ under consideration. These partial photographs must be combined into the simulated photo before a comparison with observations is possible (Section \ref{Binomial_comparison}).

\begin{table}
	\centering
	\caption{Estimated cost of updating the simulated photo by combining contributions from different processes (Section \ref{Updating_sim_photo}). This step must be protected using the $critical$ directive to avoid race conditions, and therefore cannot be parallelised. Rather, a greater number of cores would generally make this step take longer. The main parallel regions in the WBT are parallelised over $\widetilde{r}_{\text{sky}}$, so the number of steps in it is assumed to directly affect the cost of the update. Its relatively low cost suggests that the number of cores used could be increased somewhat from the nominal assumption of 60 to perhaps 120, so it may be worthwhile to use a computing cluster of this size for the WBT.}
	\begin{tabular}{cc}
		\hline
		\multicolumn{2}{c}{Update simulated photo cost ($critical$ section)} \\
		\hline
		Variable & Pixels \\
		\hline
		$\widetilde{r}_{\text{sky}}$ & 120 (average) \\
		Photo $\widetilde{v}$ & 30 \\
		Photo $r_{\text{sky}}$ & 18 \\
		Basic element (ns) & 100 \\
		Cores assumed & 1 (not parallel) \\
		\hline
		Estimated time (ms) & 6.5 \\
		\hline
	\end{tabular}
	\label{Update_sim_photo_cost}
\end{table}

The estimated cost of combining the partial photos is summarized in Table \ref{Update_sim_photo_cost}. The cost is very small, but not completely negligible as might be expected given the modest number of pixels in the photo (Tables \ref{Pixel_decision_r} and \ref{Pixel_decision_v}). This is because the simulated photo must be a shared array, raising the cost of the basic element, which involves reading from and writing to it. Another very important consideration is that updates to the simulated photo cannot be parallelised as we need to avoid race conditions between different cores, so the $critical$ directive must be used to guarantee that only one core at a time attempts to update the simulated photo. Therefore, the calculations involved in this step cannot be assumed to run in parallel across 60 cores $-$ they must run in series instead. Even so, the cost of this step is fairly small in comparison to the other steps, an important reason being that the main stages of the cost function are only parallelised over $\widetilde{r}_{\text{sky}}$, reducing the number of partial photos that must be combined to yield the simulated photo.

\subsection{Binomial comparison with observed photographs}
\label{Binomial_comparison}

To compare the resulting photograph with observations, binomial statistics will be applied to obtain $\ln P$ for each pixel when compared with each of the 10,000 pre-recorded versions of the observed photograph (Section \ref{Observations}). The total number of WB systems will in general differ between MC trials $-$ this information will be recorded in advance and used in the binomial calculations. Doing these with natural logarithms will greatly reduce the dynamic range, improving the accuracy. This has the disadvantage of raising the cost due to the need for exponentiating to get actual probabilities, but the cost of this step is sufficiently small that this is not a serious issue (Table \ref{Binomial_comparison_cost}).

The 10,000 $P$ values for each pixel will then be averaged, and the natural logarithm taken. Summing this across all pixels gives the relative log-likelihood of the model.
\begin{eqnarray}
    \label{P_calculation}
    \ln P_{\text{model}} ~&=&~ \sum_{\text{Pixels}} \ln P_{\text{pixel}} \, \text{, where} \\
    P_{\text{pixel}} ~&=&~ \frac{1}{N_{\text{trials}}} \sum_{\text{MC trials}} P \, ,
\end{eqnarray}
with $\ln P$ found using binomial statistics. To improve the accuracy with which the average $P$ is found for each pixel by considering different IP sampled photos, the maximum $\ln P$ will first be subtracted before the exponentiation step, then added back on to $\ln P_{\text{pixel}}$.

\begin{table}
	\centering
	\caption{Estimated cost of using binomial statistics to determine the absolute likelihood of each model (Section \ref{Binomial_comparison}), parallelised over the underlined variables due to the simplicity of doing so.}
	\begin{tabular}{cc}
		\hline
		\multicolumn{2}{c}{Binomial photograph comparison cost} \\
		\hline
		Variable & Pixels \\
		\hline
		\underline{$r_{\text{sky}}$} & 18 \\
		\underline{$\widetilde{v}$} & 30 \\
		Observed photos & 10000 \\
		Basic element (ns) & 150 \\
		Cores assumed & 60 \\
		\hline
		Estimated time (ms) & 1.35 \\
		\hline
	\end{tabular}
	\label{Binomial_comparison_cost}
\end{table}

\subsection{The overall cost of considering a model}
\label{Overall_cost}

Combining the stages discussed in this section leads to the overall estimated cost of evaluating each model's absolute binomial likelihood (Table \ref{Cost_function_cost}). No individual stage completely dominates the estimated cost, suggesting a reasonably optimised algorithm. The total estimated cost of 0.8s on 60 cores indicates that the WBT should be affordable with modern computer clusters. The restart protocol is fairly simple with an MCMC chain, so losses from an unplanned shutdown should be limited to only very slightly more than the time required to restart the calculation.

\begin{table}
	\centering
	\caption{Estimated time required to evaluate the cost function. The indicated tables provide a more detailed breakdown of each entry. It is possible that the actual speed will be $10\times$ slower for a long computation since the hardware might need to be protected by reducing the thermal load. In this scenario, 100,000 MC trials should take about 1 week on 60 cores. Note that for simplicity, some stages will use fewer than 60 cores, mainly if the computational cost is so low that overheads associated with further parallelisation might raise the cost further (e.g. Section \ref{Marginalising_WB_over_M}).}
	\begin{tabular}{lc}
		\hline
		\multicolumn{2}{c}{Time required for the cost function} \\
		\hline
		Stage & Time (ms) \\
		\hline
		Interpolate CB libraries in $\gamma$ (Table \ref{Outer_CB_library_gamma_interpolation_cost}) & 61 \\
		Marginalise Newtonian WB library (Table \ref{Marginalise_Newton_WB_cost}) & 18 \\
		Marginalise MOND WB library (Table \ref{Marginalise_MOND_WB_cost}) & 88 \\
		Apply stretch factors for desired $\alpha_{\text{grav}}$ (Table \ref{Stretch_factor_application_cost}) & 138 \\
		Marginalise WB library over $M$ (Table \ref{Marginalise_WB_over_M_cost}) & 23 \\
		Bunch WB libraries in $a$ (Table \ref{Bunching_WB_library_in_a_cost}) & 0.4 \\
		Get CB contamination pattern (Table \ref{CB_contamination_pattern_determination_cost}) & 188 \\
		Stretch WB $\widetilde{v}$ for hidden mass (Table \ref{Hidden_mass_stretching_cost}) & 2.4 \\
		Get velocity transformation (Table \ref{Velocity_transformation_determination_cost}) & 167 \\
		Apply velocity transformation (Table \ref{Velocity_transformation_application_cost}) & 47 \\
		Update simulated photo (Table \ref{Update_sim_photo_cost}) & 6.5 \\
		Compare with observed photos (Table \ref{Binomial_comparison_cost}) & 1.35 \\
		\hline
		Total & 753 \\
		Cores assumed (fewer in some stages) & 60 \\
		\hline
	\end{tabular}
	\label{Cost_function_cost}
\end{table}

\section{Possible extensions to the nominal analysis}
\label{Extensions}

Since the cost function is fully deterministic, gradient descent could be used to explore variations to the main analysis. If some modelling assumption has only a small impact on the best-fitting model parameters, then this is likely sufficient to demonstrate robustness to that assumption. If the impact is more substantial, the MCMC analysis should be rerun.

\subsection{The WB mass ratio prior}

Instead of a flat $q_{\text{ext}}$ distribution, some slope could be introduced. Another quite useful variation is to introduce an extra $\delta$-function in the prior on $q_{\text{int}}$ and $q_{\text{ext}}$ at exactly $1/2$ to account for the overabundance of WBs in which $q_{\text{ext}} = 1/2$ \citep{Badry_2019_twin}. This is discussed at some length in Appendix \ref{Equal_mass_CBs}.

It is also possible to alter the $q_{\text{int}}$ distribution beyond merely adding a $\delta$-function, but as the CBT (Section \ref{CB_template}) would need to be rerun for this extension, it would be very expensive.

\subsection{Varying the prior on $a_{\text{int}}$}

The CB population is expected to follow Equation \ref{P_a_CB}. In general, $P \left( a_{\text{int}} \right)$ should follow the low-end logarithmic slope used for the WBs as they are likely part of the same population. However, it is possible to vary this slope from the nominal value of $-1$ \citep{Opik_1924}. Since the CBT would need to be remade for this extension, nearly all steps would have to be repeated.

\subsection{Using line velocities}
\label{Using_line_velocities}

To limit the role of the unknown RV of each WB, it is possible to focus on $\widetilde{v}_{\text{line}}$ along the direction orthogonal to the sky position of both stars in the WB. This is similar to the line velocity method of \citet{Banik_2019_Centauri}, which argued that line velocities are much more precise, potentially avoiding systematics arising from the use of $\widetilde{\bm{v}}_{\text{sky}}$. However, the use of only one component of the proper motion doubles the number of systems needed for the WBT.

In this case, it would still be necessary to consider only systems with e.g. $\widetilde{v}_{\text{sky}} < 8$ to reduce LOS contamination. To reduce edge effects, we will consider only pixels in the observed photographs with $\widetilde{v}_{\text{line}} < 4.52$. The main advantage of the line velocity method arises from the possibility that it is known much more accurately. This means that a very restrictive upper limit of e.g. 0.1 can be imposed on the uncertainty in $\widetilde{v}_{\text{line}}$ while still retaining a significant number of systems. This could alleviate some systematics associated with the WBT, though the random errors would increase as there would be less data per WB.

We expect the technique will not prove very valuable because perspective effects from the unknown RV should be fairly small (Appendix \ref{Assigning_RV}). It might be more useful to simply repeat the main analysis with a tighter limit on the estimated uncertainty in $\widetilde{v}_{\text{sky}}$, e.g. reducing this from 0.5 to 0.25, with the reduction in sample size mitigated by allowing a larger uncertainty at high $\widetilde{v}$ beyond the critical range of $1-1.5$ upon which the WBT chiefly depends.

\section{Summary}

The WBT offers a promising way to distinguish between Newtonian and Milgromian dynamics because WBs probe the low accelerations typical of galactic outskirts, but their much smaller physical size than a galaxy means that dark matter should negligibly affect a WB even if it comprises most of the galaxy. In the Solar neighbourhood, MOND predicts that WBs should orbit each other 20\% faster than in Newtonian dynamics for an interpolating function consistent with galaxy rotation curves \citep{Banik_2018_Centauri}. Thanks to their work, the WB library is already available after performing the minor adjustments and additional calculations described in Section \ref{Further_WB_calculations}.

The main complication to the WBT is expected to be the handling of undetected companions to one or both stars in the WB (Section \ref{CB_role}). Preparing the 1CB library should be fairly inexpensive and involve a modest memory load (Section \ref{Close_binaries}). The 2CB library is likely to provide the peak computational load for the WBT, but even this should be manageable (Table \ref{Two_CB_cost}). Moreover, it should only be prepared once for the nominal analysis, even if this is extended as described in Appendix \ref{Equal_mass_CBs}.

Unlike the preparation of the CB and WB libraries, the actual analysis is certain to be repeated with different modelling assumptions (Section \ref{Extensions}), and/or with the same ones to check for dependence on the starting point. The amount of information required in the random access memory should not pose a problem (Table \ref{Master_core_memory_load_summary}).
It should take a few hours to conduct a gradient descent taking 1000 iterations with $\approx 10$ evaluations each time (one trial evaluation and one for each model parameter to obtain the gradient by finite differencing, see Table \ref{Model_parameter_list}).

This would essentially avoid the burn-in period with the MCMC stage. Even so, the large number of free parameters means this stage might require $10^5$ iterations. In this case, the whole analysis should take about a day. This would allow variations to be explored, for instance changing the prior on $q_{\text{ext}}$ and $q_{\text{int}}$, possibly in distinct ways (Section \ref{Extensions}). Some tests are required to ensure the minimum found is global, though these could be based on gradient descent starting at different points rather than on MCMC. This may well be possible because the cost function is envisaged to be fully deterministic, so it should be differentiable. It is likely that the MCMC analysis will also be more resilient with a deterministic cost function. In addition, relatively secure modelling assumptions could be tested purely by gradient descent, especially if the minimum moves very little when these assumptions are changed in some way.

Our overall assessment is that if local WBs are sufficiently well observed by Gaia and if the main source of contamination to the WBT is undetected CBs with properties that are independent of the WB (except for the upper limit to $a_{\text{int}}$), then the WBT is possible using about 60 cores simultaneously. Therefore, the WBT is expensive but probably feasible using currently available technology.

\section*{Acknowledgements}

IB is supported by Science and Technology Facilities Council grant ST/V000861/1 led by Hongsheng Zhao, the primary objective of which is to implement the WBT. IB is grateful for support from a ``Pathways to Research'' fellowship from the University of Bonn. The authors acknowledge helpful discussions with Benoit Famaey and Elena Asencio regarding this plan. They also thank Rodrigo Ibata for demonstrating operation of the 2CB preparation algorithm on the computing cluster at Strasbourg after some numerical modifications he prepared.

\bibliographystyle{mnras}
\bibliography{WBP_bbl}

\begin{appendix}

\section{Required input data}
\label{Required_input_data}

The following observables will be required for each star in a WB within 250 pc of the Sun that satisfies the quality criteria discussed in Section \ref{Observations} \citep[see also][]{Pittordis_2019}:
\begin{enumerate}
	\item position, parallax, and proper motion, and their uncertainties (5 variables $\times \, 2$),
	\item correlation coefficients between the above quantities (10 variables),
	\item Gaia-band apparent magnitude and uncertainty (1 variable $\times \, 2$),
	\item quality flags, especially the goodness of fit of the Gaia single-source astrometric solution.
\end{enumerate}
These are required for WBs with $r_{\text{sky}} = 1.5-40$ kAU and $\widetilde{v}_{\text{sky}} \leq 8$ to capture WBs that might be scattered into the photographic range in at least one MC trial (Section \ref{Observations}). Further restrictions can be applied before the main analysis based on the MC trials discussed in Section \ref{WB_MC_trials}.

\section{Restricting the MCMC analysis to physical parameter values}
\label{Restricting_MCMC_range}

Some of the model parameters (Table \ref{Model_parameter_list}) have physical restrictions on the allowed range, e.g. the slope of $P \left( e \right)$ must lie in the range $\left( -2, 2 \right)$ to avoid negative probabilities (Equation \ref{gamma_definition}). To implement such restrictions, we will repeat the part of the MCMC analysis where the proposal is set using Gaussian random numbers. Normally, only one proposal is drawn per MCMC step. However, we will draw as many proposals as are necessary to obtain one where every parameter is within its physically allowed range. The cost of obtaining a new proposal without evaluating its binomial likelihood is assumed to be negligible in comparison to the other costs, even if several hundred unphysical proposals must first be discarded at every MCMC step.

\section{Derivation of stellar mass from luminosity}
\label{Mass_luminosity_conversion}

Gaia astrometry determines the trigonometric parallax and thus the heliocentric distance of a star. Uncertainties in this will be propagated using the $5\times5$ covariance matrix, as discussed in Section \ref{WB_MC_trials}. For a known parallax, the apparent Gaia-band magnitude $G$ can be converted into an absolute magnitude $m_{_G}$, where we use the same notation as \citet{Pittordis_2019} to distinguish apparent from absolute magnitudes in different colour bands.

The stellar mass $M$ is then found from $m_{_G}$ using a Newton-Raphson algorithm based on the tabulated mass-luminosity relationship in table 5 of \citet{Pecaut_2013}, which is available at
\href{https://www.pas.rochester.edu/~emamajek/EEM\_dwarf\_UBVIJHK\_colors\_Teff.txt}{https://www.pas.rochester.edu/~emamajek/EEM\_dwarf\_UBVIJHK\_colors\_Teff.txt} with updated values. Since this is given in the $V$-band, the first step is to guess the mass and thereby obtain $m_{_V}$ and the $V - I$ colour. The latter is used to obtain the $G - V$ colour using the first Johnson-Cousins relation in table C2 of \citet{Riello_2021}. This allows the determination of $m_{_G}$, whose difference with the true value is the error of the Newton-Raphson algorithm.

In this way, we will obtain a relation between $\ln M/M_\odot$ and $m_{_G}$. We will then fit a cubic to it using similar techniques to those verified in \citet{Asencio_2021}. This cubic fit can readily be inverted using the Newton-Raphson algorithm to obtain the mass from $m_{_G}$ during each MC trial (Section \ref{WB_MC_trials}). The use of a fitting function also avoids problems of multiple masses having the same luminosity due to small-scale structure in the mass-luminosity relation, which could otherwise make an inversion very difficult.

\section{Finding WB\lowercase{s} in the Gaia catalogue}
\label{WB_discovery}

Starting with a list of $\approx 10^7$ point-like sources with reliable parallaxes $>4$~mas, we will search for pairs with a small angle on the sky. This will require 2D binning in the Galactic latitude $b$ and longitude $l$. Since the distribution of stars is close to isotropic within 250~pc, each star only needs to be compared with $\approx 10^3$ stars because the maximum allowed angular separation corresponds to an angular area $\approx 10^4 \times$ smaller than the whole sky. Assuming each pairwise comparison can be done in under 10,000 ns, the entire pair-finding operation should have a manageable cost.

If the latitude $b$ is such that the star is close to either Galactic pole, the coordinate system becomes nearly singular. Thus, a revised coordinate system will be used if $\left| b \right| > 45^\circ$. The basic idea is to use the $x$-axis as the polar direction. To keep the system right-handed and minimize the adjustments, this pole switch will be implemented using
\begin{eqnarray}
    x \leftrightarrow z \, , ~ y \leftrightarrow -y \, .
\end{eqnarray}
In the new coordinate system, the latitude is
\begin{eqnarray}
    b ~=~ \arcsin z \, .
\end{eqnarray}
To find the longitude, we will first find
\begin{eqnarray}
    r ~\equiv~ \sqrt{x^2 + y^2} \, .
\end{eqnarray}
If $r$ is very small, the longitude $l$ is irrelevant, and can be set to 0 for safety. However, this case should not arise due to the above-mentioned safety catch. As a result, $l$ will be found using a previously verified algorithm that implements
\begin{eqnarray}
    && l ~=~ \arcsin \left( \frac{y}{r} \right) \\
    && \text{if}~x < 0  \nonumber \\
    && \quad l ~=~ \mathrm{\pi} - l \\
    && \text{elseif}~y < 0 \nonumber \\
    && \quad l ~=~ l + 2\mathrm{\pi} \\
    && \text{end} \nonumber
    \label{Longitude_determination}
\end{eqnarray}

To speed up the calculations, the coordinates for every star will be stored in both systems. If a star is close to either pole in one coordinate system, then it and the few degree region around it will be close to the equator in the other coordinate system. Thus, only the coordinates in the latter system are necessary. We will exploit this by storing the coordinates of a star in a system only if the star has $\left| b \right| < 80^\circ$ in that system. Since the search region is much smaller than $35^\circ$, the binary finding algorithm can be run entirely within one or other coordinate system.

If it becomes necessary to transform not only sky positions but also proper motions, then the directions of the polar and azimuthal vectors are required in both systems. These can be defined independently of the coordinate system. For some axis $\widehat{\bm{n}}$, the polar vector is the vector on the surface of the sphere which most rapidly goes towards the pole. It is therefore
\begin{eqnarray}
    \widehat{\bm{n}}_{_b} ~\propto~ \widehat{\bm{n}} - \left( \widehat{\bm{n}} \cdot \widehat{\bm{r}} \right) \widehat{\bm{r}} \, .
\end{eqnarray}
In a right-handed system, the azimuthal vector is
\begin{eqnarray}
    \widehat{\bm{n}}_{_l} ~\propto~ \widehat{\bm{n}} \times \widehat{\bm{r}} \, .
\end{eqnarray}
In both cases, the proportionality constant is $1/\sin b$, which can be found most efficiently using $1/\sqrt{1 - \left( \widehat{\bm{n}} \cdot \widehat{\bm{r}} \right)^2}$.

Once the above-mentioned basic 2D tree structure is used to identify relevant pairs for comparison, the WB selection criteria will have to be imposed. There should be exactly one other source within the relevant 3D volume, so the parallax information must be considered as well. Since parallaxes are relatively less precise than $r_{\text{sky}}$, it will be sufficient for the two parallaxes to be consistent within the $\approx 2\sigma$ combined uncertainty. Importantly, there should be no third star close by on the sky with a parallax consistent within $\approx 3\sigma$ regardless of its proper motion, since even an unbound star passing close to a WB would still perturb it.

The precise selection criteria must be discussed with observers. We expect to consider only WBs with $\widetilde{v}_{\text{sky}} < 8$. The uncertainty on $\widetilde{v}_{\text{sky}}$ is also important $-$ we expect to only consider systems where this is $<0.5$, possibly with a tighter tolerance at low $\widetilde{v}$ (Section \ref{Using_line_velocities}).

\section{Assigning a systemic RV}
\label{Assigning_RV}

The systemic RV of a WB affects $\widetilde{v}_{\text{sky}}$ by inducing an apparent extra motion along the sky-projected separation of the WB \citep{Badry_2019, Banik_2019_Centauri}. For small angular separation WBs, the induced apparent velocity has magnitude $v_r \, r_{\text{sky}}/d$, where $v_r$ is the heliocentric RV, and $d$ is the heliocentric distance to the geometric centre of the WB. For a typical WB separated by $r_{\text{sky}} = 10$ kAU at $d = 100$~pc, this contribution is 25 m/s for $v_r = 50$~km/s, corresponding to $\Delta \widetilde{\bm{v}}_{\text{sky}} = 0.06$ along the line between the WBs if both stars are Sun-like. For systems with other properties, $\Delta \widetilde{\bm{v}}_{\text{sky}}$ could be larger. Thus, the systemic RV cannot be neglected. However, the difference in RV between the stars in a WB can safely be neglected if it is bound with a gravity law not too different from Newtonian. The possibility of LOS contamination is handled in the statistical analysis (Section \ref{Cost_function}), and is expected to be rare \citep[see Section \ref{CB_role} and][]{Pittordis_2019}.

Despite its importance, the systemic RV of each WB will generally be unknown prior to dedicated follow-up of the WBs. To estimate $v_r$, the basic principle will be to determine the relative likelihood that the WB is part of the Galactic thin or thick disk. A random number will be used to select the component, and then to assign $v_r$ based on its mean and dispersion at the position of the WB. In Galactocentric cylindrical polar coordinates, the velocity ellipsoid will be assumed constant over the survey volume as this is only a small fraction of the Sun's distance from the Galactic centre. As an example, the velocity ellipsoid tilts by $\approx 6^\circ$ per kpc in a cylindrically aligned system, so the tilt over 250 pc should be very small \citep{Hagen_2019}.

For each Galactic component, the mean $v_r$ must be found based on the assumption of the disk components rotating around the Galactic centre at a speed differing from that of the Sun by the appropriate combination of the Oort constants. The non-circular motion of the Sun also needs to be considered \citep{Francis_2014}.

The dispersion in $v_r$ for each component is
\begin{eqnarray}
	{\sigma_{_{\text{LOS}}}}^2 = \sum_{i = r,\phi,z} {\sigma_i}^2 \left( \widehat{\bm{\sigma}}_i \cdot \widehat{\bm{r}} \right)^2 \, ,
	\label{sigma_LOS}
\end{eqnarray}
where $\sigma_i$ is the velocity dispersion along Galactocentric cylindrical polar direction $\widehat{\bm{\sigma}}_i$, and $\widehat{\bm{r}}$ is the sky position of the WB's geometric centre.

\subsection{Selecting the Galactic component}

The main difficulty is in finding the appropriate membership probability $P_j$ to Galactic component $j$. We will start with a position-based prior
\begin{eqnarray}
	\rho_j ~\propto~ \rho_{j,\text{mid}} \exp \left( -\frac{\left| z \right|}{h_j} \right) \, ,
\end{eqnarray}
where $\rho_{j,\text{mid}}$ is the mid-plane density of component $j$ with exponential scale height $h_j$ in the vertical direction $z$. Other $z$-dependencies can also be considered, for instance $\sech^2$.

To find how this prior is altered by the observed $\bm{v}_t$, it is necessary to find the sky-projected velocity dispersion ellipsoid. The velocity dispersion along any direction can be found by generalising Equation \ref{sigma_LOS}. Defining two orthogonal directions within the sky plane ($\equiv$ the plane orthogonal to $\widehat{\bm{r}}$), the angle $\psi$ with respect to the first then defines any direction within the sky plane at the geometric centre of the WB. By varying $\psi$ using 1D gradient descent, we will find the direction $\widehat{\bm{\sigma}}_{\text{min}}$ along which the velocity dispersion is minimal, with value $\sigma_{\text{min}}$. The direction $\psi \pm \mathrm{\pi}/2$ is denoted $\widehat{\bm{\sigma}}_{\text{max}}$, with associated velocity dispersion $\sigma_{\text{max}}$ also found using Equation \ref{sigma_LOS}. The relative probability of Galactic component $j$ based on the observed position and $\bm{v}_t$ is then
\begin{eqnarray}
	P_j \propto \rho_j \exp \left( -\frac{1}{2} \overbrace{\sum_i \left(\frac{\Delta \bm{v}_t \cdot \widehat{\bm{\sigma}}_i}{\sigma_i}\right)^2}^{{\chi_j}^2} \right) \, ,
\end{eqnarray}
where $i$ takes two values covering the directions $\widehat{\bm{\sigma}}_{\text{min}}$ and $\widehat{\bm{\sigma}}_{\text{max}}$, while $\Delta \bm{v}_t$ is the difference between the observed $\bm{v}_t$ and the central expectation for Galactic component $j$. The Gaussian factor of $2 \mathrm{\pi} \sigma_{\text{min}} \sigma_{\text{max}}$ has been omitted from the denominator as only relative probabilities are considered.

Once results are available for both Galactic disk components, the $P_j$ will be normalised to a sum of 1. The WB will be rejected if it is inconsistent with disk kinematics in the sense that the minimum of the two ${\chi_j}^2$ exceeds some threshold enclosing e.g. 99\% of the total probability of a 2D Gaussian. A logical flag will be used to record if this is the case.

For each WB, we will need to store whether it can reliably be considered part of either Galactic disk, the two $P_j$, the mean and dispersion in $v_r$ if it belongs to Galactic disk component $j$, and the uncertainties in $\widetilde{v}_{\text{sky}}$ and $\widetilde{v}_{\text{line}}$ (Section \ref{Using_line_velocities}) for quality control purposes only. These uncertainties will be simple $1/\sqrt{N - 1}$ standard deviations found by MC error propagation (Section \ref{WB_MC_trials}). Note that since $\bm{v}_t$ varies between MC trials, the membership probabilities to different disk components must be recalculated for each trial.

\section{Inferring LOS distances}
\label{Deprojected_distances}

The method described in \citet{Banik_2019_Centauri} will be used to infer the distribution of the LOS separation of each WB, which in general is only known statistically based on the sky-projected separation, which differs slightly between MC trials. The actual separation of each WB along the LOS will be allocated using an MC approach sampling its distribution. To avoid any artificial breaks in power-law dependencies, the logarithmic slope of the $a$ distribution will be fixed at the high-end slope of the WB population as published in \citet{Banik_2018_Centauri} $-$ this should be very accurate over the vast majority of the relevant range. A broken power law can also be assumed, but for consistency this should be the same as the $P \left( a \right)$ assumed for the WB population, which is one of the model parameters to be fitted (Table \ref{Model_parameter_list}). The extra complexity means this is not envisaged in the nominal analysis.

\section{Pixel area corrections}
\label{Pixel_area_corrections}

The velocity transformation described in Section \ref{Velocity_transformation_determination} involves binning in the space of $\left( \widetilde{v}_\perp, \widetilde{v}_\parallel \right)$ to arrive at a total $\widetilde{v}$. This makes use of Pythagoras' Theorem based on the mid-point of each pixel. To improve the accuracy, we will first determine the area of each bin in total $\widetilde{v}$ in the space of $\left( \widetilde{v}_\perp, \widetilde{v}_\parallel \right)$, which involves finding the area of an annulus. This necessarily involves a factor of $\mathrm{\pi}$. We will then find the total area of the pixels in $\left( \widetilde{v}_\perp, \widetilde{v}_\parallel \right)$ assigned to the corresponding annulus, which involves simply counting pixels and multiplying by the area of each pixel. Since $\mathrm{\pi}$ is irrational, there is a small mismatch between these areas. For instance, the innermost annulus should have an area of $\mathrm{\pi}$, but instead only 3 pixels with unit area are assigned to this annulus. To approximately correct for this, we will determine the ratio between the area of the annulus and the area of the pixels assigned to it. These pixel area corrections will then be used to improve the accuracy of the algorithm. The corrections become very small at high $\widetilde{v}$, and are in any case never more than a few percent. These corrections will not be applied to the DI array described in Section \ref{Velocity_transformation_determination} since it does not rely on annular binning.

\section{CB\lowercase{s} with exactly equal mass}
\label{Equal_mass_CBs}

The WB population has an excess of systems where the stars have an exactly equal mass \citep{Badry_2019_twin}. Regardless of why this is, it should be considered in the WBT. Presumably, a similar excess is also present in the undetected CB population.

Since $\Delta \widetilde{\bm{v}} = \bm{0}$ for a WB star contaminated by an equal mass CB, we will use a special protocol based on the DI array to ensure that absolutely no velocity perturbation is applied for the induced recoil effect arising from equal mass CBs. The only variable on the CB side is therefore $\Delta \widetilde{M}$, which cannot be neglected especially in the equal mass CB case. Therefore, we will set up an array called $\bm{P}_{\text{0dv}}$ holding the likelihood of different $\Delta \widetilde{M}$ values arising from the case that $q_{\text{int}}$ is exactly $1/2$. Thanks to the 1CB case, many of the $\Delta \widetilde{M}$ bins receive a contribution as $q_{\text{ext}}$ can be $\ll 1/2$.

\subsection{Distribution of the WB mass ratio}
\label{Equal_mass_CBs_qext_prior}

To have an affordable computational cost, the resolution in $q_{\text{ext}}$ will be 0.05, so 19 steps will be required to cover the range $0.025-0.975$. Each pixel will be represented by a calculation at the central value of $q_{\text{ext}}$. Therefore, a flat prior on $q_{\text{ext}}$ would normally be imposed by setting all bins to have
\begin{eqnarray}
    P \left( q_{\text{ext}} \right) ~=~ \frac{1}{19} \, .
    \label{P_qext_no_eqm}
\end{eqnarray}
To allow an extra $\delta$-function in the distribution of $q_{\text{ext}}$ at exactly $1/2$ such that the likelihood of an exactly equal mass WB is $P_{\text{eqm}}$, Equation \ref{P_qext_no_eqm} will be adjusted to
\begin{eqnarray}
    P \left( q_{\text{ext}} \right) ~=~ \frac{1 - P_{\text{eqm}}}{19} \, + \, P_{\text{eqm}} \delta \left( q_{\text{ext}} - \frac{1}{2} \right)\, .
    \label{P_qext_with_eqm}
\end{eqnarray}
The $\delta$-function will be implemented by a boost to the probability in the central bin, which for 19 pixels is bin 10.

If some slope is introduced on the $q_{\text{ext}}$ distribution, then the idea will be for $P \left( q_{\text{ext}} \right)$ to rise linearly starting from some non-zero $q_{\text{ext}}$ until $q_{\text{ext}} = 1/2$, before decreasing symmetrically down to zero such that there is exact symmetry between $q_{\text{ext}} \to 1 - q_{\text{ext}}$. The two triangular regions defined by $P \left( q_{\text{ext}} \right)$ should have a total area of $1 - P_{\text{eqm}}$, since we also need to add an extra $\delta$-function at $1/2$. Preliminary results suggest that it will be best to assume $P_{\text{eqm}} = 0.02$ and no WBs in our sample with $q_{\text{ext}} \leq 0.125$, as a constant slope over the range $0.125-0.5$ provides a good fit.

In general, the basic principle is to combine a smooth and symmetric $q_{\text{ext}}$ distribution with a $\delta$-function at $1/2$. The smooth distribution should be integrated over the $q_{\text{ext}}$ range of each pixel to find the probability in it, with the pixel containing $q_{\text{ext}} = 1/2$ receiving a further contribution of $P_{\text{eqm}}$. Note that since we need the $q_{\text{ext}}$ distribution of Gaia WBs but the CB contamination depends on the $q_{\text{int}}$ distribution of all CBs, we expect to use different distributions for $q_{\text{int}}$ and $q_{\text{ext}}$. Relative to the underlying WB population, there should be a deficit of Gaia WBs where $q_{\text{ext}}$ differs greatly from $1/2$ because a very faint WB companion might remain undetected.

\subsection{The 1CB case}
\label{Equal_mass_CBs_1CB_library}

The $\bm{P}_{\text{0dv}}$ array is required mainly to handle the 1CB case, for which $\Delta \widetilde{M}$ depends on how much of the total detected system mass is in the contaminated star. This can be worked out by first obtaining $x$ in Equation \ref{Blended_light_factor} for $q_{\text{int}} = 1/2$, and then scaling it according to Equation \ref{q_scaling_M}. Most bins in $\bm{P}_{\text{0dv}}$ will receive a contribution once the full range of $q_{\text{ext}}$ is considered.

To improve the accuracy of this step, we will use a backward binning protocol in which we find the range of $q_{\text{ext}}$ that yields a contribution to each element of $\bm{P}_{\text{0dv}}$. By considering also the minimum and maximum allowed $q_{\text{ext}}$ (Section \ref{Equal_mass_CBs_qext_prior}), we will find the overlap range, and thus the integral of $P \left( q_{\text{ext}} \right)$ across this. A special allowance will be required in case this overlap range includes $q_{\text{ext}} = 1/2$, since then we will need to add $P_{\text{eqm}}$. The clause for this special case must arise for only one of the $\bm{P}_{\text{0dv}}$ bins, providing an important cross check. Another check is that before the MCMC stage, the $\bm{P}_{\text{0dv}}$ array should sum to 1 as factors related to $f_{\text{CB}}$ and $P_{\text{eqm}}$ will be put in as part of each MCMC trial (Appendix \ref{Equal_mass_CBs_MCMC_adjustments}).

Since part of the 1CB probability will be handled through the $\bm{P}_{\text{0dv}}$ array, the normalisation of the $\matr{P}_{\text{1CB}}$ array should be reduced before the MCMC stage. We will arrange this by implementing
\begin{eqnarray}
    \matr{P}_{\text{1CB}} ~\to~ \left( 1 - P_{\text{eqm}} \right) \matr{P}_{\text{1CB}} \, .
    \label{P_1CB_scaling_eqm}
\end{eqnarray}
It will become clear that a similar adjustment is not required for the 2CB case because a separate library will be prepared rated for the desired $P_{\text{eqm}}$ (Appendix \ref{Equal_mass_CBs_2CB_library})

Note that the pixel with $q_{\text{int}} = 1/2$ should receive some probability even if a uniform prior is assumed on $q_{\text{int}}$. As a result, the value of $P_{\text{eqm}}$ used here should exceed that in Appendix \ref{Equal_mass_CBs_qext_prior}, where the true value should be used. Equation \ref{P_1CB_scaling_eqm} should use a value which is higher by $1/999$ because the values of $q_{\text{int}}$ assumed in Table \ref{CB_template_cost} cover the range $\left(0.5-499.5\right)\perthousand$, so an extra pixel with half as much weight should be added at $q_{\text{int}} = 1/2$ if the goal is a uniform prior over the range $0.0005-0.5$. Since this argument also applies to the 2CB case and the scalings are implemented in the main algorithm, at this point we will set
\begin{eqnarray}
	P_{\text{eqm}} ~\to~ P_{\text{eqm}} + \frac{1}{999} \, .
	\label{P_eqm_adjustment_no_eqm_CBs}
\end{eqnarray}
The $\bm{P}_{\text{0dv}}$ array will then be scaled to the desired $f_{\text{CB}}$ and $P_{\text{eqm}}$ as discussed in Section \ref{Equal_mass_CBs_MCMC_adjustments}.

\subsection{The 2CB case}
\label{Equal_mass_CBs_2CB_library}

For the 2CB case, the total $\Delta \widetilde{\bm{v}}$ can only be assumed to vanish if $q_{\text{int}} = 1/2$ for both CBs. If so, only the largest $\Delta \widetilde{M}$ bin would receive a contribution. While the likelihood may seem small, a significant amount of hidden mass with no induced recoil on either WB star could mimic the effect of a gravity law stronger than Newtonian. This is one major motivation to handle the equal mass CB case with some care. In the 2CB case where both CBs have an exactly equal mass, the contribution to $\bm{P}_{\text{0dv}}$ is ${P_{\text{eqm}}}^2 {f_{\text{CB}}}^2$ into the bin with the highest $\Delta \widetilde{M}$. Since $f_{\text{CB}}$ is a model parameter (Table \ref{Model_parameter_list}), this contribution must be applied during the MCMC stage. To optimise this, ${P_{\text{eqm}}}^2$ will be determined beforehand, but after Equation \ref{P_eqm_adjustment_no_eqm_CBs} has been applied.

It is possible that only one of the two CBs is an exactly equal mass CB. To handle this case, we will set up an extension to the standard $\matr{P}_{\text{2CB}}$ library, with the extension denoted $\matr{P}_{\text{2CB,e}}$ in what follows. This will hold the CB contamination pattern for each $q_{\text{ext}}$ on the assumption that star 1 is contaminated by an exactly equal mass CB, and that it comprises a fraction $q_{\text{ext}}$ of the detected WB mass. Star 2 is assumed to be contaminated by a CB with a uniform $q_{\text{int}}$ distribution, so the CBT will be used (Section \ref{CB_template}). Since there is an asymmetry between $q_{\text{ext}} \to 1 - q_{\text{ext}}$, the library cannot be restricted to $q_{\text{ext}} \geq 1/2$, and will thus have a similar size to the 1CB library (Table \ref{One_CB_size}). The computational cost should also be similar, and thus vastly smaller than for the 2CB library (Section \ref{Two_CB_library}).

The $\matr{P}_{\text{2CB,e}}$ library will be used to create a revised $\matr{P}_{\text{2CB}}$ library where none or one of the CBs has $q_{\text{int}} = 1/2$. This is because the case where both CBs have $q_{\text{int}} = 1/2$ leads to no photocentre-barycentre offset, so we will handle it through the $\bm{P}_{\text{0dv}}$ array. Not considering this case means that the normalisation of the $\matr{P}_{\text{2CB}}$ array will drop by a fraction close to ${P_{\text{eqm}}}^2$.

The revised $\matr{P}_{\text{2CB}}$ library for each $q_{\text{ext}}$ will be found using
\begin{eqnarray}
    &&\matr{P}_{\text{2CB}} \left( q_{\text{ext}} \right) ~\to~ \left( 1 - P_{\text{eqm}} \right)^2 \matr{P}_{\text{2CB}} \left( q_{\text{ext}} \right) \\
    &&+ P_{\text{eqm}} \left( 1 - P_{\text{eqm}} \right) \left[ \matr{P}_{\text{2CB,e}} \left( q_{\text{ext}} \right) + \matr{P}_{\text{2CB,e}} \left( 1 - q_{\text{ext}} \right) \right] \, . \nonumber
\end{eqnarray}
The basic principle is that $q_{\text{ext}}$ tells us the WB mass fraction in star 1, not whether star 1 or star 2 is contaminated by an equal mass CB. Both options are possible, albeit not simultaneously $-$ as explained above, this possibility is handled later. We will safeguard the original $\matr{P}_{\text{2CB}}$ library by setting up another version rated for the desired value of $P_{\text{eqm}}$. Note that this is required even for a uniform prior on $q_{\text{int}}$ (Equation \ref{P_eqm_adjustment_no_eqm_CBs}).

\subsection{Adjustments to the main algorithm}
\label{Equal_mass_CBs_MCMC_adjustments}

In the main part of the MCMC analysis, the $\bm{P}_{\text{0dv}}$ array must be scaled to the desired $f_{\text{CB}}$ and $P_{\text{eqm}}$. A separate copy of the array will be required during the MCMC stage so that the original is still available after each MCMC trial. Since the original $\bm{P}_{\text{0dv}}$ array is entirely about the 1CB case (Section \ref{Equal_mass_CBs_1CB_library}), it must be scaled by $2 P_{\text{eqm}} f_{\text{CB}} \left( 1 - f_{\text{CB}} \right)$. To handle the 2CB case, ${P_{\text{eqm}}}^2 {f_{\text{CB}}}^2$ must then be added to the bin with the highest $\Delta \widetilde{M}$ (Appendix \ref{Equal_mass_CBs_2CB_library}).

To include probability contributions from cases with $\Delta \widetilde{\bm{v}} = \bm{0}$, the DI array will receive contributions not only from the `raw' WB library to handle the 0CB case, but also a version for every $\Delta \widetilde{M}$ where a stretch is applied in the velocity for the hidden mass effect (Equation \ref{Hidden_mass_effect}). Each time, the contribution will be scaled by the corresponding element in the $\bm{P}_{\text{0dv}}$ array. As with the nominal analysis, the DI array will then be directly injected into the simulated photo on the last $\Delta \widetilde{M}$ index, by which time the DI array must have received all the contributions it is supposed to receive. Note that the range of velocity indices must be that determined for the highest $\Delta \widetilde{M}$, since the range also scales according to Equation \ref{Hidden_mass_effect}.

\end{appendix}

\bsp
\label{lastpage}
\end{document}